\documentclass[review]{elsarticle}

\usepackage{lineno,hyperref,comment}
\modulolinenumbers[5]

\journal{Journal of Computational Physics}

\usepackage{geometry,amsmath,mathtools,amssymb}
\usepackage{ulem}
\usepackage{xcolor}
\newcommand*{\pd}[1]{\partial_{#1}}
\newcommand*{\be}{\begin{equation}}
\newcommand*{\ee}{\end{equation}}
\newcommand*{\bea}{\begin{eqnarray}}
\newcommand*{\eea}{\end{eqnarray}}
\newcommand*{\bmeq}{\begin{multiequations}}
\newcommand*{\emeq}{\end{multiequations}}
\newcommand*{\bseq}{\begin{subequations}}
\newcommand*{\eseq}{\end{subequations}}

\newcommand{\ekman}{Ek}

\newcommand{\statevector}{\mathcal{X}}
\def\avk#1{\textcolor{black}{#1}}

\def\kj#1{#1}

\def\coral{Coral } 

\def\lb{\left ( }
\def\rb{\right ) }
\def\lsq{\left [ }
\def\rsq{\right ] }
\def\lbr{\left \langle }
\def\rbr{\right \rangle }

\def\Rat{\widetilde{Ra}}
\def\Pr{Pr}
\def\ub{\mathbf{u}}
\def\vb{\mathbf{v}}

\def\dst{{\partial_t}}

\def\hz{{\bf\widehat z}}

\def\rRa{\widetilde{Ra}}

\begin{document}

\begin{frontmatter}

\title{Rescaled Equations for Well-Conditioned Direct Numerical Simulations of Rapidly Rotating Convection}

\author[boulder]{Keith Julien\footnote{Deceased on 14th April 2024.}}
\author[berkeley]{Adrian van Kan\footnote{Corresponding Author Email: avankan@berkeley.edu}}
\author[ec_lyon]{Benjamin Miquel}
\author[berkeley]{Edgar Knobloch}
\author[edinburgh]{Geoffrey Vasil}

\affiliation[boulder]{organization={Department of Applied Mathematics, University of Colorado Boulder},
            city={Boulder},
            state={Colorado 80309},
            country={USA}}

\affiliation[berkeley]{organization={Department of Physics, University of California at Berkeley},
            city={Berkeley},
            state={California 94720},
            country={USA}}

\affiliation[ec_lyon]{organization={CNRS, Ecole Centrale de Lyon, INSA Lyon, Univ. Claude Bernard Lyon 1, LMFA, UMR5509},
        city={69130 Ecully},
        country={France}}

\affiliation[edinburgh]{organization={School of Mathematics and the Maxwell Institute for Mathematical Sciences, James Clerk Maxwell Building, University of Edinburgh}, city={Edinburgh, EH9 3FD},country={UK}}

\begin{abstract}
Convection is a ubiquitous process driving geophysical/astrophysical fluid flows, which are typically strongly constrained by planetary rotation on large scales. A celebrated model of such flows, rapidly rotating Rayleigh-Bénard convection, has been extensively studied in direct numerical simulations (DNS) and laboratory experiments, but the parameter values attainable by state-of-the-art methods are limited to moderately rapid rotation (Ekman numbers $\ekman\gtrsim10^{-8}$), while realistic geophysical/astrophysical $\ekman$ are significantly smaller. Asymptotically reduced equations of motion, the nonhydrostatic quasi-geostrophic equations (NHQGE), describing the flow evolution in the limit $\ekman\to 0$, do not apply at finite rotation rates. The geophysical/astrophysical regime of small but finite $\ekman$ therefore remains currently inaccessible. Here, we introduce a new, numerically advantageous formulation of the Navier-Stokes-Boussinesq equations informed by the scalings valid for $\ekman\to0$, the \avk{\textit{Rescaled Rapidly Rotating incompressible Navier-Stokes Equations} (RRRiNSE)}. We solve the \avk{RRRiNSE} using a spectral quasi-inverse method resulting in a sparse, fast algorithm to perform efficient DNS in this previously unattainable parameter regime.  We validate our results against the literature across a range of $\ekman$, and demonstrate that the algorithmic approaches taken remain accurate and numerically stable at $\ekman$ as low as $10^{-15}$. Like the NHQGE, the \avk{RRRiNSE} derive their efficiency from adequate conditioning, eliminating spurious growing modes that otherwise induce numerical instabilities at small $\ekman$. We show that the time derivative of the mean temperature is inconsequential for accurately determining the Nusselt number in the stationary state, significantly reducing the required simulation time, and demonstrate that full DNS using \avk{RRRiNSE} agree with the NHQGE at very small $\ekman$.
\end{abstract}


\end{frontmatter}

\nolinenumbers
\section{Introduction}
\label{sec:intro}
Buoyant convection in the presence of rotation represents a ubiquitous scenario for geophysical and astrophysical fluid flows that is largely responsible for the turbulent dynamics observed in planetary and stellar interiors \cite{jones2011planetary,roberts2013genesis,miesch2000coupling}, and in planetary atmospheres \cite{ingersoll1990atmospheric,emanuel1994atmospheric} and oceans \cite{marshall1999open,soderlund2019ocean}. The dynamics are highly complex
with many influential ingredients such as geometry, compressibility, multiple components, and the presence of magnetic fields. In the absence of such complexities, the quintessential paradigm for investigating rotationally influenced buoyant flows is provided by rotating Rayleigh-B\'enard convection (RRBC). A large number of studies has been published on this model system, which is very well suited for detailed experimental, numerical and theoretical studies, including \cite{chandrasekhar1953instability,nakagawa1955theoretical,veronis1959cellular,rossby1969study,pfotenhauer1984stability,boubnov1986experimental,zhong1991asymmetric,boubnov2012convection,julien1996rapidly,knobloch1998rotating,hart2002mean,vorobieff2002turbulent,stevens2013heat},
to name but a few. In its most distilled form, the problem consists of a rotating plane layer of fluid confined between two parallel horizontal plates which maintain a destabilizing temperature gradient. However, the interpretation of a \textit{layer} within this paradigm may be broadened to include confined fluid domains such as cylinders, annuli, as well as spherical interiors and shells, which often arise in geophysical and astrophysical applications.

Five nondimensional parameters of geophysical and astrophysical interest highlight the relative importance of the Coriolis, pressure gradient, buoyancy, and dissipation forces in setting the acceleration of the fluid.
These are the bulk Rossby, Euler, buoyancy, Reynolds and P\'eclet numbers, respectively:
\be
\label{eq:nond}
Ro_H = \frac{U}{2\Omega H}, \quad
Eu = \frac{P}{\rho_0 U^2},\quad
\Gamma_H = \frac{g\alpha\lVert \nabla T_b \rVert H^2}{U^2},\quad
Re_H = \frac{UH}{\nu},\quad
Pe_H = \frac{UH}{\kappa}=\Pr Re_H,
\ee
which are comprised of intrinsic, extrinsic and characteristic properties of the fluid. Intrinsic material properties include the coefficient of thermal expansion $\alpha$, the kinematic viscosity $\nu$, and the thermal diffusivity $\kappa$, with $\Pr=\nu/\kappa$ denoting the Prandtl number. Extrinsic properties include the magnitude $\Omega$ of the rotation rate, the layer depth $H$, the gravitational acceleration $g$ and the applied temperature gradient $\lVert \nabla T_b \rVert$. Characteristic properties include the velocity $U$, pressure $P$, and the constant reference density $\rho_0$. The subscript `$H$' signifies association with the bulk layer depth. Also of importance is the ratio of the viscous and Coriolis forces that provides an \textit{a priori} external parameter referred to as the Ekman number 
\be
\label{eqn:Ek}
\ekman = \frac{Ro_H}{Re_H} = \frac{\nu}{2\Omega H^2}.
\ee
\begin{table}
\begin{center}
\begin{tabular}{|l|c|c|c|c|}
\hline
Celestial body & $\ekman$ & $\Pr$  & $Ro_H$ & $Re_H$ \\
\hline
Earth's outer core & $10^{-15}$ & $0.1$ &  $10^{-7}$ & $10^8$ \\ 
Mercury (core) &  $10^{-12}$ & $0.1$  & $10^{-4}$ & $10^8$ \\ 
Jupiter (core) & $10^{-19}$ & $0.1$  & $10^{-10}$ & $10^9$ \\ 
\ \ \ Europa (ocean) & $10^{-12}$ & $11.0$ & $10^{-2.5}$--$10^{-1.5}$ & $10^{9.5}$--$10^{10.5}$ \\ 
\ \ \ Ganymede (ocean) & $10^{-10}$--$10^{-13}$ & $10.0$  & $10^{-3.5}$--$10^{1.5}$ & $10^{9.5}$--$10^{11.5}$ \\ 
Saturn (core) & $10^{-18}$ & 0.1  & $10^{-9}$ & $10^9$ \\
\ \ \  Enceladus (ocean) & $10^{-10}$--$10^{-11}$ & $13.0$  & $10^{-3.5}$--$10^{-1}$ & $10^{7.5}$--$10^{9}$ \\ 
\ \ \ Titan (ocean) & $10^{-11}$--$10^{-12}$ & $10.0$  & $10^{-3}$--$1$ & $10^{9}$--$10^{11}$ \\ 
Neptune (core) & $10^{-16}$ & 10.0  & $10^{-6}$ & $10^{10}$ \\
Uranus (core) & $10^{-16}$ & 10.0  & $10^{-6}$ & $10^{10}$ \\
\hline
\end{tabular}
\end{center}
\caption{Nondimensional parameter estimates for planetary \cite{schubert2011} and satellite interiors \cite{soderlund2019ocean}. 
Estimates of the Rossby number are derived from the relation $Ro_H=Re_H Ek$.}
\label{Table:Paramo}
\end{table}

Turbulent flows are characterized by $Re_H\gg1$, and when strongly influenced by rotation, by \kj{the ordering} $\ekman\ll Ro_H \lesssim 1$ (cf. Eq.~\eqref{eqn:Ek}). 
Table I provides estimates of these nondimensional parameters in important geophysical and astrophysical settings. It can be seen that all such flows are rapidly rotating ($\ekman\ll1$), highly turbulent ($Re_H\gg 1$), and in the majority of situations strongly influenced by rotation ($Ro_H \ll 1$).  To first approximation, using the theory of isotropic and statistically stationary turbulence as a benchmark,
an order of magnitude estimate of the range of scales between the integral and dissipative scales in terms of the number of degrees of freedom per spatial direction and time is given, respectively, by $Re_H^{3/4}\ge \mathcal{O}(10^6)$ and $Re_H^{1/2}\ge \mathcal{O}(10^4)$ \citep{sP00,frisch1995turbulence}.  Probing this region of parameter space is further complicated by an extended temporal range. Specifically, the smallness of the Rossby number $10^{-10} \le Ro_H \le 10^{-1}$  indicates an extreme timescale separation between fast inertial waves, associated with the Coriolis force, propagating on $\mathcal{O}(\Omega^{-1})$ time scales and the motion of eddies evolving on the advective time scale $\mathcal{O}(H/U)$. From the combined values of $Re_H$ and $Ro_H$, Table \ref{Table:Paramo} indicates that this extended temporal range may span as much as ten logarithmic decades.
\begin{figure}[h]
    \centering
    \includegraphics[width=0.8\textwidth]{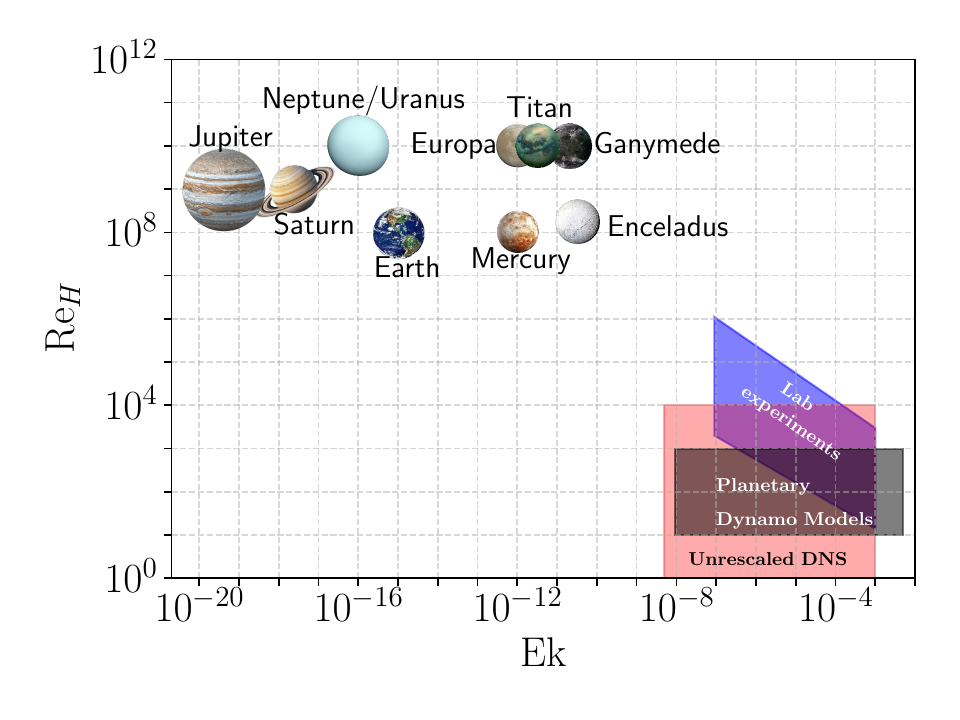}
    \caption{Overview of the parameter space of RRBC spanned by the Ekman number $\ekman$ and the bulk Reynolds number $Re_H$. 
    Experiments, simulations, and dynamo models populate the parameter space characterized by moderately large Ekman numbers and moderately low Reynolds numbers (shaded) but these are far from their geo-/astrophysically relevant values. Adapted from \cite{aurnou2015rotating}, based on Table~\ref{Table:Paramo}. }
    \label{fig:Param_Re_vs_Ek}
\end{figure}

From the standpoint of direct numerical simulations (DNS) -- which are required to resolve all scales of the motion -- these estimates are truly daunting. \textcolor{black}{The current capability of state-of-the-art 3D DNS is $\mathcal{O}(10^4)$ degrees of freedom in each spatial direction for periodic boundary conditions \cite{kaneda2006high, yeung2015extreme,yeung2025gpu} 
indicating an upper threshold of $Re_H=\mathcal{O}(10^5)$. In the presence of walls, the current state of the art in 3D DNS is $\mathcal{O}(10^3)$ degrees of freedom in each spatial direction \cite{blass2021effect,howland2024turbulent,song2024direct}, further restricting the accessible range of $Re_H$.} 
Hence, directly accessing the geophysical and astrophysical parameter regime is out of reach for the foreseeable future, even with impending advances to exascale supercomputing. Figure~\ref{fig:Param_Re_vs_Ek} captures this void visually. Recent DNS studies place the threshold at $\ekman\gtrsim 10^{-8}$, $Re_H\lesssim 10^4$, e.g. \cite{king2012heat,favier2014inverse,guervilly2014large,kunnen2016transition,favier2020robust,guzman2021force,song2024scaling,song2024direct}. Inclusion of spherical geometry and the capability for dynamo action in the simulations further restricts the reported range to $\ekman\gtrsim 10^{-7}$, $Re_H\lesssim 10^3$ \cite{cooper2020subcritical,mason2022magnetoconvection,kolhey2022influence,majumder2024self}, although 
\avk{hyperviscous simulations have been used to emulate lower Ekman numbers \cite{aubert2017spherical}.}
Given these limitations, a popular DNS strategy has been to vary $Re$ over the accessible range and attempt to uncover scaling laws in global quantities such as momentum and heat transport with a goal of extrapolating the results to the geophysical and astrophysical settings of Table~\ref{Table:Paramo}. 
However, to be physically meaningful, such an extrapolation must be performed while respecting the strong rotational constraint $Ro_H=Re_H \ekman \ll1$. Inspection of this expression indicates that this gives rise to the challenging and somewhat incompatible requirement that the Ekman number be repeatedly lowered as $Re_H$ increases (see also Fig.~\ref{fig:Param_Re_vs_Ek}). This leads to an amplification in the stiffness of the governing equations due to an increased separation between the time scales of inertial waves and advection, as well as between the advective and dissipation time scales (see Section~\ref{sec:spatemp}). As a result, this requirement imposes severe time-stepping constraints on the majority of numerical algorithms currently in use. 

\kj{The issue resides, in particular, with the precise implementation of the time-stepping scheme.} 
Specifically, the linear Coriolis force $2\Omega U\hz\times\ub$, of relative order $Ro_H^{-1}$ compared to inertial forces, is often treated explicitly, e.g. in \cite{julien1996rapidly,guervilly2014large,verzicco1996, king2010convective,mP16,zhu2018}, while the advective timescale associated with the nonlinear advection term, $\ub\cdot\nabla\ub$, is invariably treated explicitly and so is known from a prior time step. Algorithmically, this avoids the complexities of implementing a coupled numerical solver for the momentum equations, and by contrast permits the use of a decoupled solver that updates fluid variables sequentially at each time step. However, several recent codes for simulating rapidly rotating convection, including \cite{pM16,burns2020dedalus, bM21}, treat the Coriolis force implicitly. This formulation has a number of advantages and we use it below to identify a rescaled RRBC model and numerical algorithms capable of accessing regimes characterized by $Re_H\gg1$ and $\ekman \ll Ro_H \ll1$. 

The remainder of this paper is organized as follows. \avk{In Section~\ref{sec:iNSE}, the incompressible Navier-Stokes equations (iNSE) in the Boussinesq approximation are given in a rotating frame.} In Section~\ref{sec:spatemp}, the detailed spatiotemporal resolution requirements for buoyantly driven, rotationally constrained flows are discussed and the need for implicit time-stepping treatments is highlighted.  In Section~\ref{sec:asymp_guide}, an asymptotically reduced set of equations, the nonhydrostatic quasi-geostrophic equations (NHQGE), is established as an instrumental guide for deducing a reformulation of the full iNSE. 
Informed by the asymptotic equations, Section~\ref{sec:rescaled_dns} introduces a novel formulation of the iNSE termed the \avk{rescaled rapidly rotating incompressible Navier-Stokes equations (RRRiNSE)}. Section \ref{sec:cond} highlights some of the advantageous numerical properties of \avk{RRRiNSE}, establishing that the numerical discretization is well conditioned. Section \ref{sec:results} contains a detailed comparison of fully nonlinear DNS using the newly introduced reformulation with established results from the literature, together with an analysis of the mean temperature equation, along with novel DNS results for the Nusselt and Reynolds numbers at unprecedented Ekman numbers ($\ekman$ as low as $10^{-15}$ and smaller). Finally, Section \ref{sec:conclusions} concludes with a discussion of the implications of our results for future numerical simulations of rapidly rotating convection. Where necessary, relevant detailed calculations are relegated to  
Appendices.

\section{The incompressible Navier-Stokes Equations: iNSE}
\label{sec:iNSE}

In the classic paradigm of rotating Rayleigh-B\'enard convection in a horizontal plane layer the fluid motion is accurately captured by the Boussinesq approximation that assumes small density fluctuations about a static background state, resulting in the incompressible Navier-Stokes equations (iNSE)
\begin{subequations}
\label{eq:standard_NS_formulation_dimensional}
\bea
\label{eqn:v}
\lb \dst + \ub\cdot \nabla\rb\ub  &=& - 2\Omega\,\hz\times\ub -\nabla \pi + g\alpha\vartheta \, \boldsymbol{\hz} + \nu \nabla^2 \ub,\\
\label{eqn:c}
\nabla\cdot\ub &=& 0,\\
\label{eqn:tt}
\lb \dst + \ub\cdot \nabla\rb \vartheta + \ub\cdot\nabla T_b  &=& \kappa \nabla^2 \vartheta,
\eea
\end{subequations}
where $\ub$ represents the convective fluid velocity, $\pi$ is the modified pressure, and $T=T_b(z) + \vartheta(\boldsymbol{x},t)$, i.e., the temperature is split into a static background profile $T_b(z)$ in the vertical direction and a convective temperature contribution $\vartheta$. The system rotates at a constant frequency $\Omega$ about the vertical  direction $\hz$; the rotational Froude number is assumed to be sufficiently small that the centrifugal force can be neglected. 

The equations of motion can be nondimensionalized by a characteristic but as yet undetermined flow velocity scale $U$, the layer depth $H$, and the characteristic temperature gradient $\lVert \nabla T_b \rVert$ giving
\begin{subequations}
\label{eq:standard_NS_formulation_dimless}
\bea
\label{eqn:v}
\lb \dst + \ub\cdot \nabla\rb\ub  &=& - \frac{1}{Ro_H} \hz\times\ub - Eu \nabla \pi + \Gamma_H \vartheta \boldsymbol{\hz} + \frac{1}{Re_H}  \nabla^2 \ub,\\
\label{eqn:c}
\nabla\cdot\ub &=& 0,\\
\label{eqn:tt}
\lb \dst + \ub\cdot \nabla\rb \vartheta + \ub\cdot\nabla T_b  &=& \frac{1}{Pe_H} \nabla^2 \vartheta.
\eea
\end{subequations}
The nondimensional parameters are defined in \eqref{eq:nond}. In the next section, we describe the challenges associated with solving the above set of nondimensional equations numerically in the regime $Ro_H\ll 1$, $Re_H\gg 1$.

\section{Spatiotemporal resolution requirements for buoyantly driven flow}

\label{sec:spatemp}
 A review of the rotating convection literature illustrates why the $Ro_H\ll1$ regime has proven to be so challenging for DNS \citep{aurnou2015rotating,song2024scaling, song2024direct,kunnen2021geostrophic, ecke2023turbulent}.
Hereafter, for simplicity of exposition, we focus our discussion on the case where the rotation axis is antiparallel with gravity, i.e., the polar regime where $\nabla {T_b} \equiv \partial_z {T_b} {\bf \hat{z}}$. 
 For fixed temperature boundary conditions $\partial_z {T_b}=(T_b(H)-T_b(0))/H$, while for a fixed heat flux $F$, $\partial_z T_b = - F/\kappa$ instead. Within this regime a dynamical balance exists between the ageostrophic Coriolis, inertial and Archimedean (buoyancy) forces. This so-called CIA balance \citep{jmA20} establishes the  rotational free-fall velocity
\be
U_{r\!f\!f} = \frac{g\alpha \lVert \partial_z T_b  \rVert H}{2\Omega} \equiv  Ro_c U_{\!f\!f} 
\ee
as \kj{an appropriate estimate} for the characteristic velocity, an estimate that has been verified both numerically \citep{sM21, tO23} and experimentally \citep{Hawkins2023,madonia2023,abbate2023rotating}. 

We define the convective Rossby number $Ro_c$ as the ratio of the rotational free-fall velocity to the buoyancy free-fall velocity $U_{\!f\!f}$ observed in rotationally unaffected regimes,
\be
U_{\!f\!f} = \sqrt{g\alpha \lVert\partial_z T_b\rVert H^2}\,.
\ee
Thus
\be
Ro_c \equiv \frac{U_{r\!f\!f}}{U_{\!f\!f}} = \sqrt{\frac{Ra}{Pr}}\ekman.
\ee
Here
\be
Ra = \frac{g\alpha\lVert \partial_z T_b \rVert H^4 }{\nu\kappa}
\ee
is the thermal Rayleigh number. 
This definition of $Ro_c$ is physically more precise than the {\it equivalent} definition $Ro_c=U_{\!f\!f}/(2\Omega H)$, i.e. as the Rossby number $Ro_H$ based on $U_{\!f\!f}$.

The convective Rossby number provides an {\it external} measure of the rotational constraint based on the imposed thermal Rayleigh number. Since $Ro_c\ll1$ for rotationally constrained flows, it follows that $U_{r\!f\!f}\ll U_{\!f\!f}$.

With $U=U_{r\!f\!f}$ as the correct characteristic velocity scale it follows from \eqref{eq:nond} that
\be
\label{eqn:dist}
\Gamma_H=\frac{1}{Ro^2_c},\quad
Re_H = \frac{Ro^2_c}{\ekman }.
\ee
Moreover, asymptotic linear theory \citep{chandrasekhar1953instability} and 
simulations \citep{mS06,kJ12} both indicate that rotating flows are highly anisotropic with
\be
\label{eqn:xyz}
\nabla_\perp\sim \frac{H}{\ell} \sim  \ekman^{-1/3}, \qquad \partial_Z \sim 1\,,
\ee
and an ${\cal O}(\ekman^{-4/3})$ onset Rayleigh number, indicating that horizontal variations occur on the scale $\ell\ll H$ while vertical variations occur on the scale of the layer depth. Together with the corresponding nonlinear theory \citep{kJ98,JK_PoF_1999,kJ07} these results lead to the introduction of the reduced Rayleigh number  
\be
\label{eqn:dist2}
\rRa = Ra \ekman^{4/3}\,,
\ee
with $\rRa = {\cal O}(1)$ defining the strongly forced ($Ra\gg 1$) but still rotationally constrained ($Ro_c\ll 1$) regime of interest.
This regime extends from the convective threshold to highly supercritical Rayleigh numbers subject to the requirement that $\rRa$ is no larger than $\rRa = o(\ekman^{-1/3})$. This upper bound represents the constraint required to maintain the local rotational constraint  
\be
Ro_\ell \equiv \frac{U_{r\!f\!f}}{2\Omega \ell} =  Ro_c^2 \frac{H}{\ell}\sim Ro_c^2\ekman^{-1/3} =o(1).
\ee 

It is now clear that the external order parameter threaded throughout the rapidly rotating regime is $\ekman^{1/3}$. Anticipating the derivation and discussion of the rescaled formulation, this observation suggests the definition of the small parameter:
\be
\epsilon = \ekman^{1/3}\,. \label{eq:defEpsilon}
\ee

\begin{table}[t]
\begin{center}
\begin{tabular}{|c|c|c||c|c|c|c|}
\hline
 CFL time step             & rotation  & horiz. diff.   &horiz. adv.   &  buoyancy       & vert. adv.  & vert. diff.        \\
 $\Delta t_{\boldsymbol{f}}$   & $\Delta t_{\Omega}$   & $\Delta t_{\nu_\perp}$  & $\Delta t_{adv_\perp}$ & $\Delta t_g$  &  $\Delta t_{adv_\parallel}$   & $\Delta t_{\nu_\parallel}$ \\ 
\hline
&                                                    &                                                          &                                                          &                                                                                         &                                                     & \\
Dimensional                                          &  $\displaystyle{\frac{1}{2\Omega}}$   & $\displaystyle{\frac{\left( \Delta x_\perp^*\right )^2}{\nu}}$ & $\displaystyle{\frac{\Delta x_\perp^{*}}{U_{r\!f\!f}}}$ & $\displaystyle{\frac{U_{r\!f\!f}}{g\alpha \theta}}$  &    $\displaystyle{\frac{\Delta z^{*}}{U_{r\!f\!f}}}$  & $\displaystyle{\frac{\left( \Delta z^*\right )^2}{\nu}}$ \\
&                                                    &                                                          &                                                          &                                                                                         &                                                     &  \\ \hline 
&                                                    &                                                          &                                                          &                                                                                         &                                                     & \\
Nondim., $\displaystyle{\frac{H^2}{\nu}}$    & $\displaystyle{\ekman}$ & $\displaystyle{\ekman^{2/3} \left ( \Delta x_\perp \right )^2}$ & $\displaystyle{\frac{\ekman^{4/3}}{Ro^2_c}\Delta x_\perp }$   & $\displaystyle{ \ekman^{2/3} }$ &  $\displaystyle{\frac{\ekman}{Ro^2_c}\Delta z}$  & $\displaystyle{\left( \Delta z\right )^2}$                                           \\
&                                                    &                                                          &                                                          &                                                                                         &                                                     &\\ \hline
&                                                    &                                                          &                                                          &                                                                                         &                                                     & \\
Nondim., $\displaystyle{\frac{\ell^2}{\nu}}$  & $\displaystyle{\ekman}^{1/3}$  &    $\displaystyle{\left( \Delta x_\perp\right )^2}$    &       $\displaystyle{\frac{\ekman^{2/3}}{Ro^2_c} \Delta x_\perp }$         &  $1$           & $\displaystyle{\frac{\ekman^{1/3}}{Ro^2_c} \Delta z}$         & $\displaystyle{\ekman^{-2/3}\left( \Delta z\right )^2}$   \\
&                                                    &                                                          &                                                          &                                                                                        &                                                     & \\ 
\hline
\end{tabular}
\end{center}
\caption{Ordering of the most to least restrictive CFL conditions for the explicit time step $\Delta t_{\boldsymbol{f}}$ associated with the forcing term $\boldsymbol{f}$ in the incompressible Navier-Stokes Equation (iNSE): 
$\partial_t \ub = \boldsymbol{f}\equiv - 2\Omega \hz\times\ub -\nabla p + \nu \nabla^2_\perp \ub - \ub\cdot\nabla_\perp\ub - g\alpha \theta \boldsymbol{\hat r} - \ub\cdot\nabla_\parallel\ub + \nu \nabla_\parallel^2 \ub $.
The time constraint for the pressure force is identical to that of the Coriolis force. Row 2 gives the dimensional time step estimate. Later rows express the nondimensional estimates based on vertical ($H^2/\nu$) and horizontal ($\ell^2/\nu$) diffusion times. Here $\ell\sim \ekman^{1/3} H$ such that $\Delta x_\perp^* = \ell \Delta x_\perp$, $\Delta z^* = H \Delta z$ where $\Delta x_\perp \propto N_{x_\perp}^{-1}$ and $\Delta z \propto N_z^{-1}$. From \citep{kJ12,jmA20}, $U_{r\!f\!f}\sim Ro_c U_{\!f\!f}$ and \avk{the temperature fluctuation $\theta$, defined in Eq.~(\ref{eq:decomposition_vartheta}), scales as $\theta\sim \ekman^{1/3} \lVert \partial_z T_b \rVert H$, where $Ro_c=\sqrt{Ra/Pr}\ekman$.} Forces in need of an implicit treatment are presented to the left of the vertical separator $\lVert $.
 \textcolor{black}{This holds provided $\Delta x_\perp \ll \rRa^{-1}$, otherwise, no advantages arise from the implicit treatment of horizontal dissipation (column 3) given that its CFL constraint becomes as restrictive as nonlinear horizontal advection (column 4).}}
\label{Table:CFL} 
\end{table}%
For given (nondimensional) grid resolutions $\Delta x$, $\Delta z\ll 1$ and temporal resolution $\Delta t\ll 1$, the assumption of an explicit time-stepping algorithm leads in Table \ref{Table:CFL} to the spatio-temporal constraints known as the Courant-Friedrich-Lewy (CFL) criteria required for accurately discretizing the various forces in the iNSE. The most to least restrictive CFL conditions are shown in dimensional (row 2) and nondimensional forms according to the vertical or horizontal diffusion timescale (rows 3 \& 4). This \kj{ordering holds provided $\Delta x\sim \Delta z < \rRa^{-1}$}. 
In the limit $\ekman\rightarrow 0$, it is clear from Table~\ref{Table:CFL} that the Coriolis term (column 2) imposes the most restrictive constraint on the time step $\Delta t$ \textcolor{black}{(provided $\Delta x\sim \Delta z \gg  \rRa E^{1/3}/\Pr$)}. This suggests that it is numerically advantageous to treat this linear term implicitly with the additional expense of numerically coupling the momentum equations. It is also evident that, \textcolor{black}{compared to nonlinear horizontal advection,} an implicit treatment is desirable for the linear horizontal dissipation, the next most prohibitive constraint, $\propto (\Delta x_\perp)^2$ \textcolor{black}{provided $\Delta x_\perp =o\lb\rRa^{-1}\rb$}. If this strategy is adopted then all remaining time-stepping bounds for the linear terms are less severe than the $\mathcal{O}(\Delta x_\perp)$ nonlinear horizontal advection timescale.
Thus all remaining linear terms can be treated explicitly without a numerical penalty. 

The mechanical conditions at an impenetrable boundary also result in additional resolution constraints in space. Specifically, no-slip boundaries and/or stress-free boundaries that are not perpendicularly aligned to the axis of rotation result in $\mathcal{O}(\ekman^{1/2})$ Ekman boundary layers. For no-slip boundaries, it has recently been established that this prohibitive constraint can be relaxed by parameterizing its effect on the bulk through the pumping boundary conditions $w = \pm \ekman^{1/2} \, \hz\cdot  \nabla\times\ub \, /\sqrt{2}$ \citep{mP16,kJ16}. \textcolor{black}{This complication does not arise in the present work, since we focus exclusively on stress-free top and bottom boundaries.}

Given the enormous challenges faced by DNS in the $Ro\ll 1$ regime, an attractive alternative is to resort to large-eddy simulations (LES), which resolve only the large turbulent scales, and employ subgrid-scale models for the smaller turbulent scales below a certain threshold scale. This technique has been applied in the context of nonrotating Rayleigh-B\'enard convection \cite{peng2006large,foroozani2018large,samuel2022large}. However, it must be stressed that, for the highly anisotropic turbulent flows encountered in the geophysical and astrophysical context, LES are still in their infancy and ill-understood due to the complex structure across scales which such flows exhibit. Even when LES can be applied, the results thus obtained still need to be extrapolated to the extreme parameter regimes of geophysical and astrophysical flows. Importantly, LES and subgrid-scale modelling are particularly challenging because there is a notable paucity of validation data in the relevant regimes.

\section{Asymptotically reduced model as a guide}
\label{sec:asymp_guide}

Attempts to increase the achievable Reynolds number in DNS (or LES) of RRBC while lowering the Ekman number to sustain the low Rossby number environment must result from improving the conditioning of the matrices obtained from numerical discretization. Ultimately this  means reducing or removing the discretization dependence on the Rossby or Ekman number. The asymptotic system of equations for RRBC valid in the limit $\ekman\to 0$ derived and extensively studied by Julien \& coworkers serves as a template for accomplishing this task \citep{mS06,  kJ12,kJ07,kJ98a}. Assuming a local plane layer about the North pole, the system leverages $Ro_c\sim \ekman^{1/3}{\equiv \epsilon}\ll1$ as the small parameter, along with the characteristic anisotropic scalings \eqref{eqn:xyz}, and the relations (\ref{eqn:dist}) and (\ref{eqn:dist2}). A primary geostrophic balance is obtained together with horizontal incompressibility on horizontal spatial scales, namely,
\be
\hz\times\ub \approx -\nabla \pi, \qquad 
\nabla_\perp\cdot\ub_\perp \approx 0.
\ee
 It follows that $\ub_\perp = \hz\cdot\nabla_\perp\times \pi \equiv 
 \nabla^\perp \pi$ where $\nabla^\perp =(-\partial_y,\partial_x)$.
Moreover, it is found that the modified pressure $\pi=\Psi$ serves as the geostrophic streamfunction with  $\ub=(\nabla^\perp \Psi, w)$.  When observed on the characteristic anisotropic spatial scales $\ell$ and $H$, and velocity scale $\nu/\ell$,  the reduced system of equations (the Non-Hydrostatic Quasi-Geostrophic Equations \citep{mS06}) is given by 
\begin{subequations}\label{eqns:reduced}
\bea
\label{eqn:va}
 \dst \nabla^2_\perp \Psi + J\lsq \Psi, \nabla^2_\perp \Psi \rsq -\partial_Z w  =  \nabla_\perp^4 \Psi,\\
\label{eqn:wa}
 \dst w  + J\lsq \Psi, w  \rsq   + \partial_Z \Psi = \frac{\Rat}{Pr} \theta  + \nabla^2_\perp w,\\
\label{eqn:fta}
 \dst \theta + J\lsq \Psi, \theta \rsq   + w  \left ( \partial_Z \overline{\Theta} - 1\right ) = \frac{1}{Pr} \nabla_\perp^2 \theta,\\
\label{eqn:mta}
\epsilon^{-2} \dst\overline{\Theta} +\partial_Z \lb \overline{w  \theta}\rb =  \frac{1}{Pr} \partial_{ZZ} \overline{\Theta}
\eea
with
\be
\label{eqn:rbc}
w = \overline{\Theta} = 0, \quad\mbox{on} \quad Z = 0,1.
\ee
\end{subequations}
Here $J\lsq \Psi, f \rsq = \ub_\perp \cdot\nabla_\perp f$,
$\hz\cdot\boldsymbol{\omega}=\nabla^2_\perp \Psi$ is the vertical vorticity and corrections at ${\cal O}({\ekman}^{1/3})$ have been dropped; $Z$ is the ${\cal O}(1)$ vertical scale. The temperature field is decomposed into a mean (horizontally-averaged) and $\mathcal{O}(\epsilon)$ fluctuating component, i.e., 
\be
\vartheta \equiv  \overline{\Theta} + \epsilon \theta. \label{eq:decomposition_vartheta}
\ee
It follows that the Ekman number dependence remains only in the evolution of the mean temperature $\overline{\Theta}$ which can be seen to evolve on a much slower timescale $T= \epsilon^{2}t$ \avk{(ratio of vertical viscous diffusion time to horizontal diffusion time)} compared to the convective dynamics.
Importantly, it has been established in \cite{mS06,kJ12,kJ98a} that this term can be omitted provided: (i) $\overline{\Theta}$ evolves to a statistically stationary state $\partial_t \overline{\Theta}\approx 0$, and (ii) the fluid domain is sufficiently large that numerous convective cells or plumes contribute to the horizontal spatial averaging. 

The result serves as an accurate representation for $\overline{\Theta}$ with $\mathcal{O}(\ekman^{2/3})$ error (see \ref{sec:tempomit}). 
Assuming an implicit treatment in time for all linear terms, it can now be seen that the most restrictive condition is the Ekman number-independent CFL constraint on the horizontal advection terms, i.e. $\Delta t = \Delta x /\lVert U_{RMS} \rVert $, which is consistent with our discussion in Table \ref{Table:CFL}. 

The NHQGE \eqref{eqn:va}-\eqref{eqn:rbc} have been instrumental in probing and identifying the properties of turbulent Rayleigh-B\'enard convection in the rapidly rotating regime, from the identification of regimes of distinct flow morphologies \citep{kJ12}, to the understanding of a novel inverse energy cascade in three dimensions \citep{sM21,aR14}, through to uncovering the dissipation-free momentum and heat transport scaling laws \citep{sM21,tO23,kJ12b}. However, by design, the NHQGE are constructed to be valid in the asymptotic regime $\ekman\ll Ro_H \rightarrow 0$\textcolor{black}{, and a complete understanding of its robustness to finite $\ekman$ remains an open question \citep{sS14}}. Bridging the intermediate void in parameter space between the regimes obtained in current laboratory experiments and DNS and actual geophysical and astrophysical settings as highlighted in Fig.~\ref{fig:Param_Re_vs_Ek} is a key scientific objective.

\section{Rescaled incompressible DNS}
\label{sec:rescaled_dns}

Based on the discussion in the previous section of the asymptotically reduced governing equations, we can now reformulate the iNSE in a rescaled form, which is advantageous for simulating very low Ekman and Rossby numbers. For this we will follow the template that produced the reduced system (\ref{eqns:reduced}). We begin by introducing the anisotropic characteristic length scales: we nondimensionalize vertical lengths by $H$ and horizontal lengths by $\ell = \epsilon H$, where $\epsilon = \ekman^{1/3}$ as before. We also adopt the velocity scale $U=U_\nu =\nu/\ell$ which differs from a rotational free-fall velocity scale according to $U_\nu = (\rRa/Pr) U_{r\!f\!f}$. This implies that 
\bea
\label{eq:distl1}
& \displaystyle{Ro_H \equiv\frac{U}{(2\Omega H)} = \ekman^{2/3}\equiv \epsilon^2,\quad 
\Gamma_H = \frac{1}{\epsilon^2} \frac{\widetilde{Ra}}{Pr},\quad 
Re_H =\frac{1}{\epsilon},\quad Pe_H=Pr Re_H=\frac{Pr}{\epsilon},}\\ 
\label{eq:distl2}
& \displaystyle{ Eu\equiv\frac{P}{\rho_0 U^2} = \frac{2 \Omega U \ell^3}{\nu^2}= \frac{1}{\ekman^{1/3}}= \frac{1}{\epsilon}}. 
\eea
Note that due to the anisotropic rescaling one finds
\begin{equation}
 \nabla_\perp \mapsto \frac{1}{\epsilon}\nabla_\perp,\hspace{0.5cm}  
\hat{\bf z} \cdot \nabla =  \partial_Z,\hspace{0.5cm} 
\partial_t \mapsto {\frac{1}{\epsilon}}\partial_t.
\end{equation}
As in the derivation of the reduced equations, we decompose the temperature deviation from the linear conductive background state according to  $\vartheta = \overline{\Theta}(Z,t) + \epsilon\theta(\mathbf{x},t)$. Finally, we define the ageostrophic velocities
\begin{subequations}
\label{eq:rinse_tot}
\begin{equation}
U = \frac{1}{\epsilon} (u + \partial_y \pi), \hspace{0.5cm} V = \frac{1}{\epsilon} (v - \partial_x \pi) \quad \Longleftrightarrow \quad \boldsymbol{U}_\perp =\frac{1}{\epsilon} (\ub_\perp - \nabla^\perp \pi) . \label{eq:ageo_uv} 
\end{equation}
Recall $\nabla^\perp = (-\partial_y, \partial_x)$ and $\nabla_\perp = (\partial_x, \partial_y)$. 
The iNSE then take the form 
\begin{align}
\dst {u} +\mathcal{N}_\epsilon u -V  = &\   \widetilde{\nabla}^2_\epsilon u. \label{eq:rinse_ux}\\
\dst v  +\mathcal{N}_\epsilon v  + U = &\ \widetilde{\nabla}^2_\epsilon v,\\
\dst w  + \mathcal{N}_\epsilon w +   \partial_Z \pi = &\ \widetilde{\nabla}^2_\epsilon w +\frac{\widetilde{Ra}}{Pr} \theta, \\
\partial_x U + \partial_y V +  \partial_Z w = &\ 0,\\
\dst \theta  + \mathcal{N}_\epsilon \theta + ( \partial_Z \overline{\Theta} - 1) w = &\ \frac{1}{Pr}\widetilde{\nabla}^2_\epsilon \theta,\\
\epsilon^{-2} \dst \overline{\Theta} +  \partial_Z  \overline{w\theta}= &\ \frac{1 }{Pr}  \partial_Z^2 \overline{\Theta}, \label{eq:rinse_tm}
\end{align}
\end{subequations}
where
\begin{equation}
 \widetilde{\nabla}^2_\epsilon = \nabla^2_\perp + \epsilon^2 \partial_Z^2, \,\qquad \nabla^2_\perp=\partial_x^2 + \partial_y^2, \quad \quad \quad \mathcal{N}_\epsilon =  u\partial_x  + v \partial_y  + \epsilon w \partial_Z .
 \end{equation}
We consider impenetrable stress-free, fixed-temperature boundary conditions, i.e., 
\begin{equation}
w=\partial_Z u = \partial_Z v = 0 \text{ at } Z=0,1,
\end{equation}
and 
\begin{equation}
\theta = \overline{\Theta} = 0 \text{ at } Z=0,1.
\label{eq:bctemp}
\end{equation}

 The equation set (\ref{eq:ageo_uv})-(\ref{eq:bctemp}) is an equivalent reformulation of the iNSE obtained by rescaling terms (without omitting any terms in the process) in accordance with the asymptotic theory, specifically utilizing the distinguished limits (\ref{eq:distl1})-(\ref{eq:distl2}) described in the previous section.  We refer to these equations as the Rescaled Rapidly Rotating incompressible Navier-Stokes Equations (\avk{RRRiNSE}). 
 
 We complete our exposition of these equations by noting that the \avk{formal} limit $\epsilon \to 0$ of these equations leads directly to the asymptotic reduced equations (\ref{eqn:va})-(\ref{eqn:mta}) describing quasi-geostrophic Rayleigh--B\'enard convection. This follows on noting that
 \begin{equation}
 \lim_{\epsilon\rightarrow 0} \widetilde{\nabla}^2_\epsilon = \nabla^2_\perp\,;
\end{equation}
moreover, introducing the streamfunction $\Psi$ and velocity potential $\chi$ decomposition of the horizontal velocity field,
\begin{equation}
    \ub_\perp = \nabla^\perp \Psi + \epsilon \nabla_\perp \chi, 
\end{equation}
where $\nabla_\perp = (\partial_x,\partial_y,0)^T$ and $\nabla^\perp = (-\partial_y,\partial_x,0)^T$, we see that
\be
\lim_{\epsilon \to 0} \pi = \Psi; \quad
\lim_{\epsilon \to 0} \boldsymbol{U}_\perp = \nabla_\perp \chi; \quad \lim_{\epsilon \to 0} \ub_\perp = \nabla^\perp \Psi\,.
\ee
Thus, as in the asymptotic equations, the dependence of the pressure on the velocity changes from quadratic to linear and a leading order geostrophic velocity field is recovered. Three-dimensional incompressibility is maintained through the ageostrophic velocity, $\nabla_\perp\cdot\boldsymbol{U}_\perp = \nabla^2_\perp \boldsymbol{\chi} = -\partial_Z w$. In Section \ref{sec:results} we  demonstrate empirically that in sufficiently large domains $\epsilon^{-2}\partial_t \overline{\Theta}\approx 0$ in the statistically steady state, including at very small values of $\ekman$.

\section{Conditioning properties of rescaled equations: \avk{RRRiNSE}}
\label{sec:cond}

The advantage of the \avk{RRRiNSE} formulation can be displayed through the properties of its spatio-temporal discretization. The findings of Table~\ref{Table:CFL} suggest an implicit-explicit time discretization scheme for the \textcolor{black}{governing equations} of the form
\be
\label{eq:tstep}
\lb \partial_t \mathcal{M} - \mathcal{L}_I \rb \statevector^{(n+1)} =  \mathcal{L}_E \statevector^{(n)} + \mathcal{N} (\statevector^{(n)},\statevector^{(n)}) 
\ee
with implicit and explicit vectors of state $\statevector^{(n+1)}\equiv(\ub^{(n+1)},\boldsymbol{U}^{(n+1)}_\perp,\pi^{(n+1)},\theta^{(n+1)})^T$ and similarly for $\statevector^{(n)}$. The exact list of the variables that enter $\statevector$ and the expressions for differential operators $\mathcal{M},\mathcal{L}_I, \mathcal{L}_E$ and $\mathcal{N}$ all depend on the adopted formulation. The specific details for various forms of \avk{RRRiNSE} and the asymptotic model NHQGE are relegated to \ref{sec:IMEX}.

Numerically, given that $\mathcal{L}_I$ is a non-diagonal operator, this requires the utilization of a coupled solver at each time step. The non-diagonal component is associated with the Coriolis force that would impose an $\mathcal{O}(\epsilon^{-1})$ explicit time-stepping constraint. The linear operator $\mathcal{L}_E$ represents vertical diffusion that, consistent with Table~\ref{Table:CFL}, imposes a nonrestrictive $\mathcal{O}(\epsilon^{-2}(\Delta z)^2)$ explicit time-stepping constraint. However, we note an apparent conflict in pursuing an explicit treatment of vertical diffusion, a term via which mechanical boundary conditions at the top and bottom 
are imposed, be it grid-based or via a basis of special functions. Two possibilities arise as resolutions to this predicament: (i) a reversion to an implicit treatment of $\mathcal{L}_E$, or (ii)
the construction of Galerkin basis functions that automatically impose the mechanical boundary constraint. Implementation of the latter is insensitive to implicit/explicit treatments while the former requires a near-boundary resolution
\be
\epsilon^{-2} \left(\Delta z_b\right )^2 \lesssim \Delta t_{adv_\perp} \propto \frac{\Delta x}{\lVert U_{\perp}\rVert_\infty}\quad\implies\quad
\Delta z_b \lesssim \epsilon \left (\Delta t_{adv\perp}\right)^{1/2} 
\ee
to ensure that the numerical scheme is aware of the boundary constraint. This is more prohibitive than the $\mathcal{O}(\epsilon^{3/2})$ resolution constraint of an Ekman boundary layer in the limit $\epsilon\rightarrow 0$. We therefore pursue
a Galerkin function approach \textcolor{black}{and, in addition, adopt an implicit treatment of $\mathcal{L}_E$, unless specified otherwise}. \avk{Parameterized Ekman pumping boundary conditions  \cite{mP16}, while not considered here, go beyond the Galerkin basis approach but can be handled via the use of the tau method \cite{ortiz1969tau} implemented in widely available solvers such as Dedalus \cite{burns2020dedalus}. }

Equations \eqref{eq:tstep} 
are solved with the numerical code \coral \citep{bM21}, a flexible platform for solving systems of PDEs with spectral accuracy, i.e., with exponential error convergence. All fluid variables are discretized with a Fourier mode expansion in the horizontal and a Chebyshev-Galerkin polynomial expansion in the vertical direction, i.e.,
\be
\label{eq:decomp}
\vb = \sum \hat{\vb}_{\boldsymbol{k}_\perp,j} (t) e^{i \boldsymbol{k}_\perp\cdot \boldsymbol{x}_\perp} \Phi_j \left( Z \right ).
\ee
Here the Chebyshev-Galerkin basis functions $\Phi_j \left( Z \right )$ \avk{can be} constructed to satisfy Dirichlet ($\Phi_j =0$), Neumann ($\partial_Z \Phi_j=0$) or stress-free conditions ($\Phi_j =\partial_{ZZ} \Phi_j = 0$) or mixtures thereof on the boundaries; \avk{here, we consider the latter case only.} The code temporally evolves the spectral coefficients of the modes $\hat{\vb}_{\boldsymbol{k}_\perp,j} (t)$ in spectral space, here via the third order-four stage implicit-explicit Runge-Kutta time-stepping scheme RK443 \citep{Ascher97}. For constant-coefficient differential equations, as considered here, \coral adopts the quasi-inverse method presented in \citep{kJ09}, based on an integral formulation of the problem, applied to Chebyshev-Galerkin bases obtained by basis recombination \citep{kJ09,burns2020dedalus,bM21}. This procedure, detailed in \ref{appendix:coral}, is implemented in \coral and results in a sparse banded structure for the coupling matrices $\mathcal{M}$ and $\mathcal{L}_I$ in \eqref{eq:tstep}. 

We note that, given the dependence of the time step $\Delta t$ on $\ekman$ through its presence in $\mathcal{L}_I$, 
the ability to take the limit ${\ekman \rightarrow}\ 0$ is ultimately bounded by the accuracy of the time integration due to the specific time-discretization error associated with the scheme and round-off errors. Such errors are ultimately related to the condition number of the matrix  $\boldsymbol{A}$ in the linear algebraic system $\boldsymbol{A}\vb=\boldsymbol{b}$ that results from the spatial-temporal discretization of \eqref{eq:tstep}. Here $\boldsymbol{A}$ is the discretization of $\partial_t \mathcal{M} - \mathcal{L}_I$ and $\boldsymbol{b}$ is the explicit right-hand side of \eqref{eq:tstep}.
Alternatively, the sensitivity of an implicit time-stepping scheme can be explored through the eigenspectrum of the generalized eigenproblem
$\lambda \boldsymbol{M} - \boldsymbol{L}_I$ deduced from the discretization of $(\partial_t \mathcal{M} - \mathcal{L}_I)\vb =0$.
\begin{figure}
\centering
\includegraphics[width=1\textwidth]{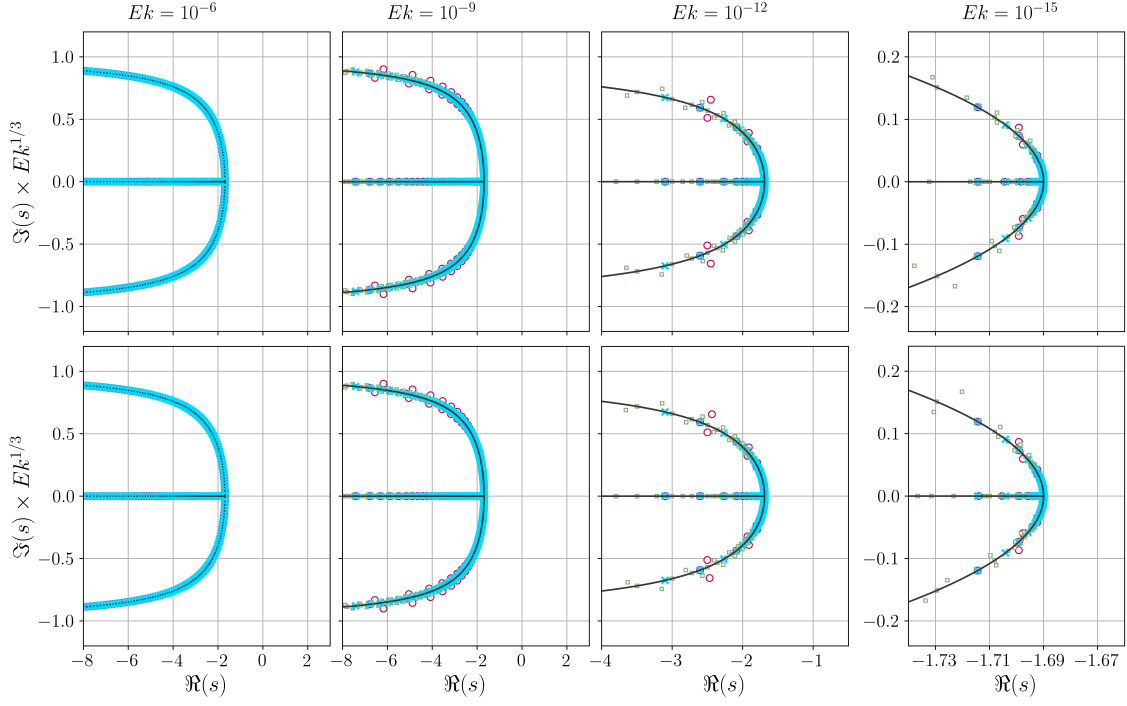}
\caption{\label{fig:rinse_spectra}Linear spectra from the rescaled equations (\ref{eq:rinse_tot}) and (\ref{eq:rinse_mixedVelocityVorticity}) in the complex plane obtained numerically using the quasi-inverse method with Chebyshev-Galerkin basis (see appendix B). The top row illustrates the case of pure inertial waves ($\widetilde{Ra}=0$) while thermal stratification below the convective onset ($\widetilde{Ra}=5$) is included in the bottom row. All panels use $\widetilde{k}_\perp=1.3$ and $Pr=1$. Numerical solutions are obtained for both the primitive variable formulation (\ref{eq:rinse_tot}) with $N_Z=256$ (magenta circles) and $N_Z=512$ (green squares), and the mixed velocity-vorticity formulation (\ref{eq:rinse_mixedVelocityVorticity}) with $N_Z=256$ (blue crosses). For reference, the analytical dispersion relation is represented with black dots appearing as a continuous black line. In both cases the computation of the numerical spectra remains stable as $Ek$ reaches values as low as $10^{-15}$. The mixed velocity-vorticity formulation leads to a remarkably accurate numerical spectrum, at the cost of larger memory usage.
}
\end{figure}
\begin{figure}
\centering
\includegraphics[width=1\textwidth]{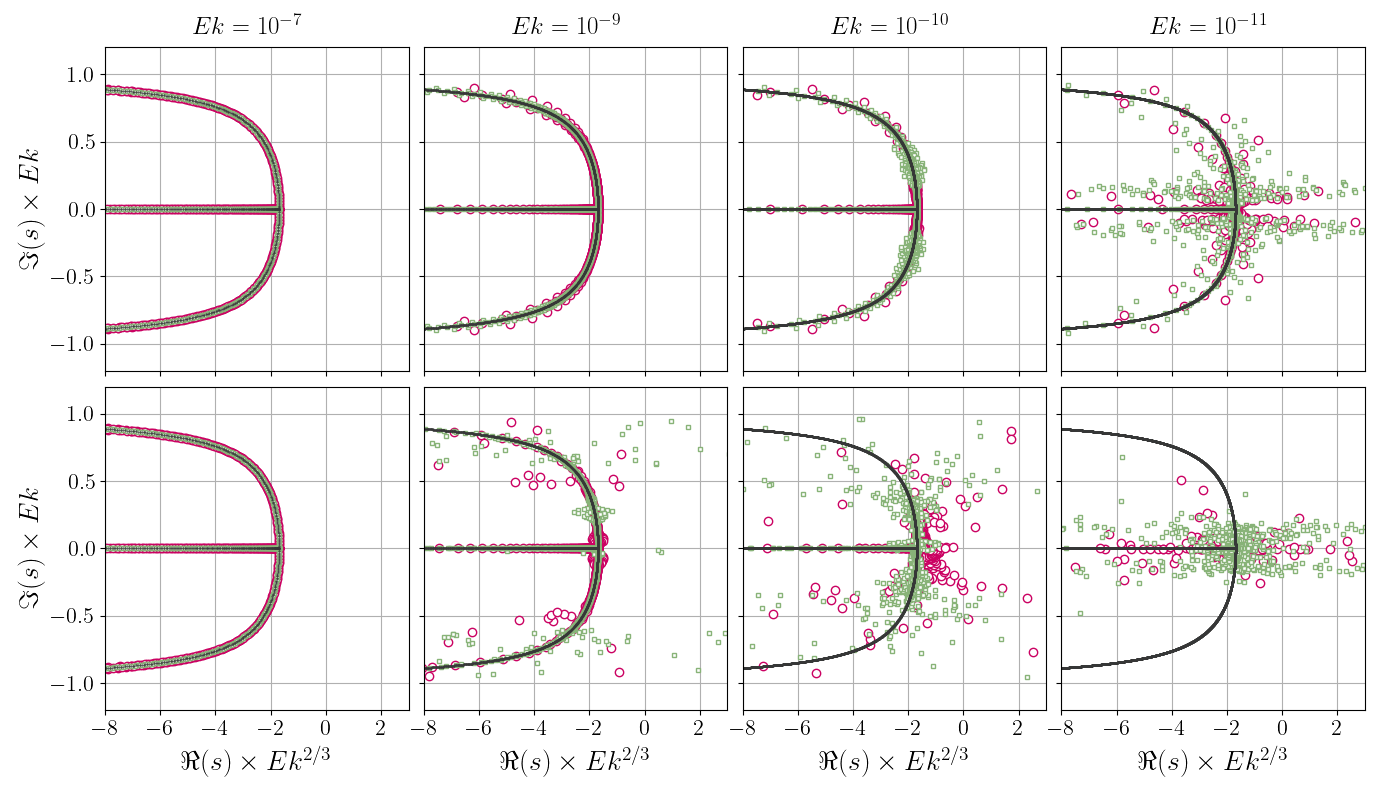}
\caption{\label{fig:dns_spectra} Linear spectra from the unscaled equations (\ref{eq:standard_NS_formulation_dimless}) in the complex plane obtained numerically using the quasi-inverse method with Chebyshev-Galerkin basis (see appendix B). The top row illustrates the case of pure inertial waves ($\widetilde{Ra}=0$) while thermal stratification below the convective onset ($\widetilde{Ra}=5$) is included in the bottom row. All panels use $\widetilde{k}_\perp=1.3$ and $Pr=1$. Numerical solutions obtained with $N_Z=256$ (magenta circles) and $N_Z=512$ (green squares) are compared against the analytical dispersion relation (black dots appearing as a continuous black line). In both cases spurious unstable modes proliferate with decreasing $Ek$.}
\end{figure}

Figure~\ref{fig:rinse_spectra} shows this eigenspectrum, obtained with the \avk{RRRiNSE} formulation, in the complex plane for four different Ekman numbers $\ekman=10^{-6},10^{-9}, 10^{-12}, 10^{-15}$ at $\widetilde{Ra}=0$ (top row) and $\widetilde{Ra}=5$ (bottom row), with the numerical results indicated by blue crosses (for the mixed velocity-vorticity formulation (\ref{eq:rinse_mixedVelocityVorticity}) with $N_Z=256$ Chebyshev modes in the vertical), magenta circles and green squares (for the primitive-variable form (\ref{eq:rinse_tot}) with $N_Z=256$ and $N_Z=512$ Chebyshev modes, respectively), in excellent agreement with the analytical result shown by the black line. This result should be compared with Fig.~\ref{fig:dns_spectra}, which displays the eigenspectrum obtained with
the unscaled Boussinesq equations at Ekman numbers $\ekman=10^{-7}, 10^{-9}, 10^{-10}, 10^{-11}$, again at $\widetilde{Ra}=0$ (top row) and $\widetilde{Ra}=5$ (bottom row), with magenta circles ($N_Z=256$) and green squares ($N_Z=512$) indicating the numerical data for comparison with the analytical result shown by the black line. In contrast with Fig.~\ref{fig:rinse_spectra}, Fig.~\ref{fig:dns_spectra} shows that the accuracy of the numerical spectra deteriorates significantly as the Ekman number decreases below $\ekman \lesssim 10^{-7}$. Particularly damaging to time-stepping the solution are the spurious, linearly unstable modes (i.e., modes with $(\Re(s)>0)$) visible in Fig.~\ref{fig:dns_spectra} for $\widetilde{Ra}=0$ when $\ekman=10^{-11}$ (top row) and even more spectacularly for the thermally forced case $\widetilde{Ra}=5$ when $\ekman=10^{-9}$ (bottom row), i.e., at a substantially slower rotation rate than in the purely hydrodynamical case. These results indicate why traditional DNS has proved unable to reach $\ekman=10^{-9}$.

\begin{figure}
\centering
\includegraphics[trim={0 0 6cm 18cm},clip, width=0.49\textwidth]{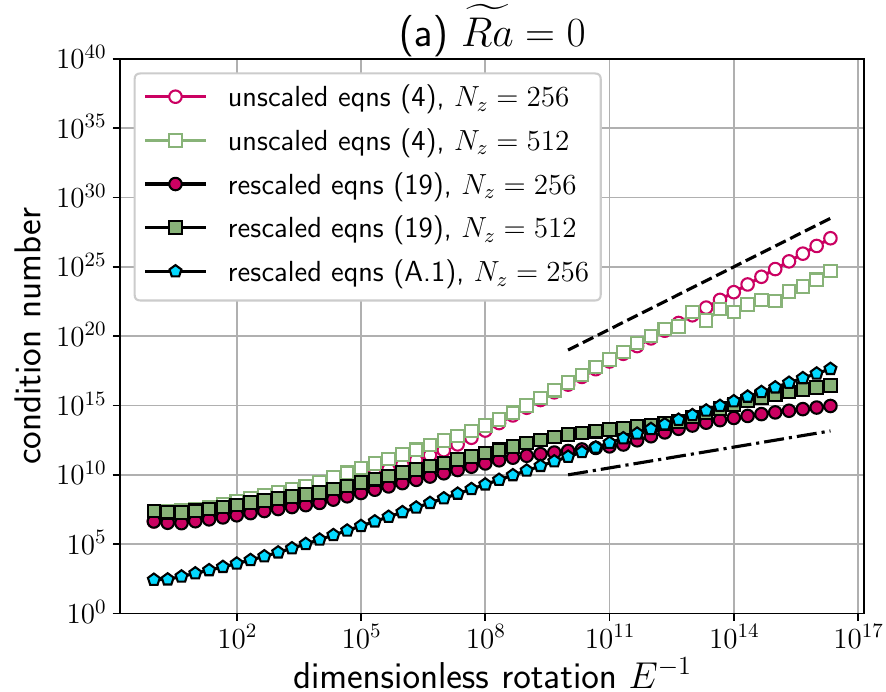}
\includegraphics[trim={0 0 6cm 18cm},clip,width=0.49\textwidth]{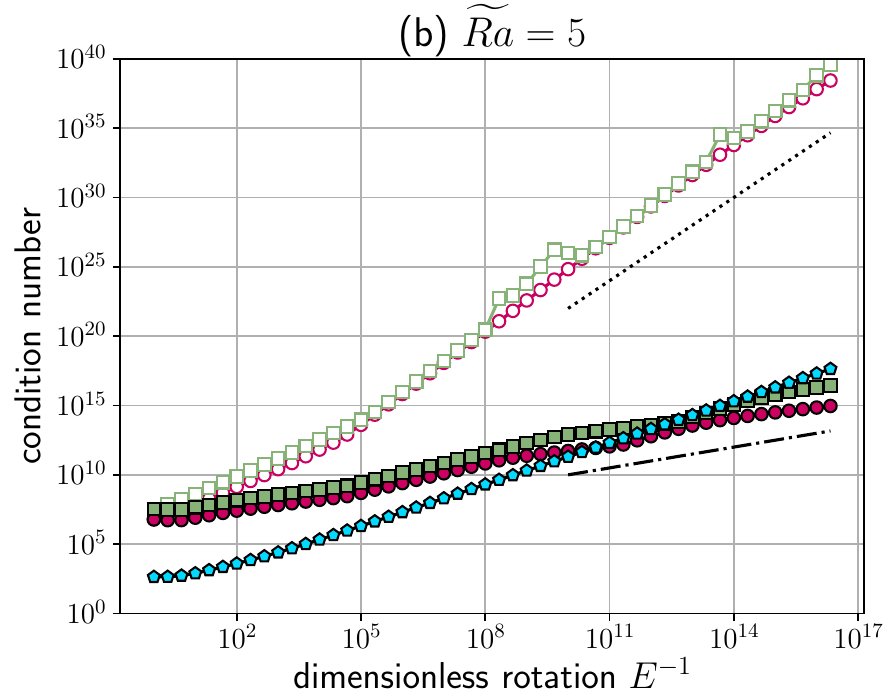}
\caption{\label{fig:condition_number} Condition number of the operator $\boldsymbol{L}_I$ computed for both the standard formulation [equation (\ref{eq:standard_NS_formulation_dimless}), open symbols] and the rescaled formulations [equations (\ref{eq:rinse_tot}) and (\ref{eq:rinse_mixedVelocityVorticity}), filled symbols]. Both panels use $\widetilde{k}_\perp=1.3$ and $Pr=1$. (a) Pure inertial waves for $\widetilde{Ra}=0$. (b) Thermally stratified case $\widetilde{Ra}=5$. The dash-dotted, dashed, and dotted lines are eye-guides with slope $1/2$, $3/2$ and $2$, respectively. }
\end{figure}

The behavior of the spectra associated with the standard and the rescaled formulations is quantified and summarized in Fig.~\ref{fig:condition_number} which presents the condition number of the operator $\boldsymbol{L}_I$ as rotation increases. High values for the condition number of matrices are commonly associated with unstable numerical computations. With the standard formulation (\ref{eq:standard_NS_formulation_dimless}), the conditioning of the discretized operator degrades rapidly as the Ekman number decreases following an approximate $\ekman^{-3/2}$ scaling law at $\widetilde{Ra}=0$, and an even steeper scaling close to $\ekman^{-2}$ at $\widetilde{Ra}=5$, as $\ekman\rightarrow 0$. For comparison, the condition number associated to the \avk{RRRiNSE} obeys a more moderate $\ekman^{-1/2}$ scaling for the primitive variable form (\ref{eq:rinse_tot}), and only a somewhat steeper scaling for the mixed velocity-vorticity form (\ref{eq:rinse_mixedVelocityVorticity}). The relative values of the condition number also speak clearly in favor of the rescaled formulation: the condition number computed with the \avk{RRRiNSE} (in either the primitive variable or mixed velocity-vorticity forms) for geophysically relevant rotation strengths ($\ekman = 10^{-16}$) appears to be smaller than its counterpart computed with the standard formulation even at modest rotation rate ($\ekman\approx 10^{-6}$). 

These results reflect the \avk{well-conditioned nature} and the practicability of the \avk{RRRiNSE} equations in the small $\ekman$ limit, which the unscaled equations do not possess. This fact provides a strong motivation for using the \avk{RRRiNSE} system to perform accurate DNS in the limit of small $\ekman$.

\section{Results}
\label{sec:results}
To further validate the \avk{RRRiNSE} formulation, going beyond the improved conditioning properties of \avk{RRRiNSE} presented in Section~\ref{sec:cond}, we perform extensive direct numerical simulations of RRBC using Coral. All runs described below were performed with $Pr=1$ and a rescaled (nondimensional) domain with dimensions $10\ell_c\times 10\ell_c\times 1$ was used throughout, \textcolor{black}{with critical convective onset length scale} $\ell_c\approx4.82$, unless specified otherwise. An explicit treatment is used for all advective \avk{RRRiNSE} terms $\mathcal{N}_i$, $i=u,v,w,\theta,\Theta$ and also for the mean temperature advection term $(\partial_Z\overline{\Theta} -1)w$ in the $\theta$ equation. The CFL condition is imposed based on the horizontal velocity components. Stress-free boundary conditions are adopted in all runs.

First, in Sections~\ref{ssec:dtheta_mean_dt} and \ref{ssec:imp_exp_vert_deriv_diffusion} we discuss different numerical schemes that can be used for solving the \avk{RRRiNSE}. These differ in the treatment of the mean temperature equation and vertical derivatives in the diffusion terms. \kj{Our results focus on the global heat and momentum transport as defined by the nondimensional Nusselt and Reynolds numbers, namely,}
\be
Nu = 1 + Pr \langle \overline{w \theta} \rangle_{Z,t} \equiv 1 + \partial_{Z} \overline{\Theta}\vert_{0,1},\qquad
Re_w = \langle\langle \overline{w^2} \rangle^{1/2}_{Z}\rangle_t\,,
\label{eq:def_Nu_Re}
\ee
where $\langle \cdot \rangle_{Z,t}$ denotes \avk{the combined average in the vertical and in time, while $\langle \cdot \rangle_Z$ and $\langle \cdot \rangle_t$ denote averaging in the vertical or in time separately.} {Along with the Nusselt number $Nu$, the quantity $Re_w$ saturates significantly earlier than the horizontal velocity components that are strongly impacted by an inverse kinetic energy cascade \citep{sM21,tO23}.} In Section~\ref{ssec:comparison_published_Nu}, we compare the Nusselt number obtained in our simulation to published results in the literature. Next, in Section~\ref{ssec:convergence_geostrophic}, we present the Nusselt and Reynolds numbers for different values of $\ekman$ and $\widetilde{Ra}$, verifying the convergence of \avk{the RRRiNSE} to the asymptotically reduced equations presented in Section~\ref{sec:asymp_guide}. Finally, we provide visualizations of our \avk{RRRiNSE} simulation results at a very low Ekman number, $\ekman=10^{-15}$, well within the geostrophic turbulence regime.

\subsection{Slaving: the role of $\epsilon^{-2}\partial_t \overline{\Theta}$ in the mean temperature equation}
\label{ssec:dtheta_mean_dt}
\begin{figure}
   \centering
   \includegraphics[width=0.32\textwidth]{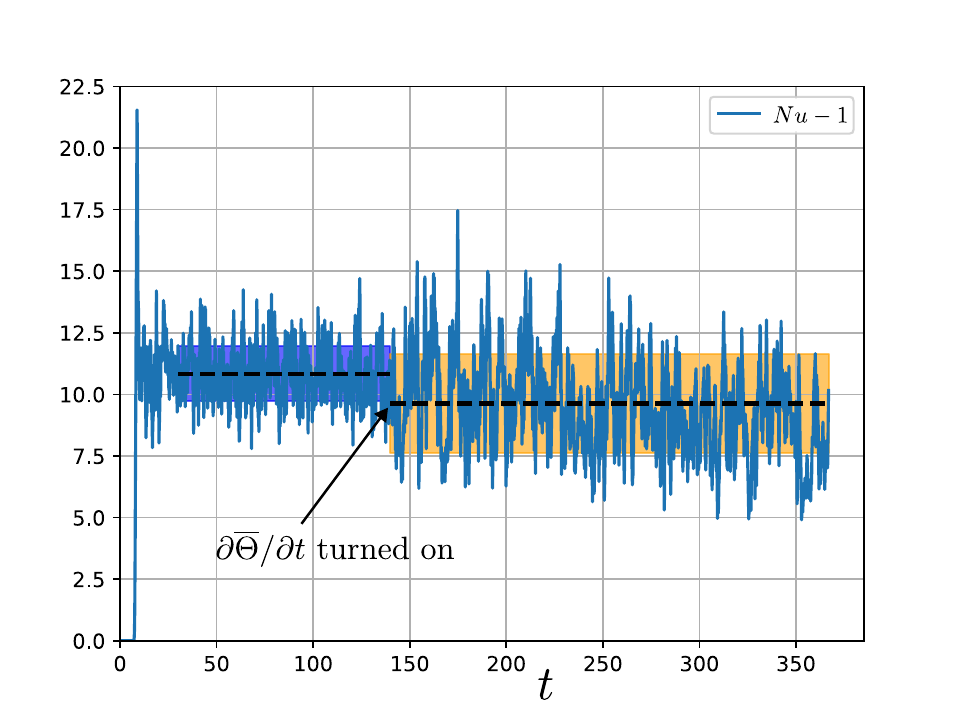}
   \includegraphics[width=0.32\textwidth]{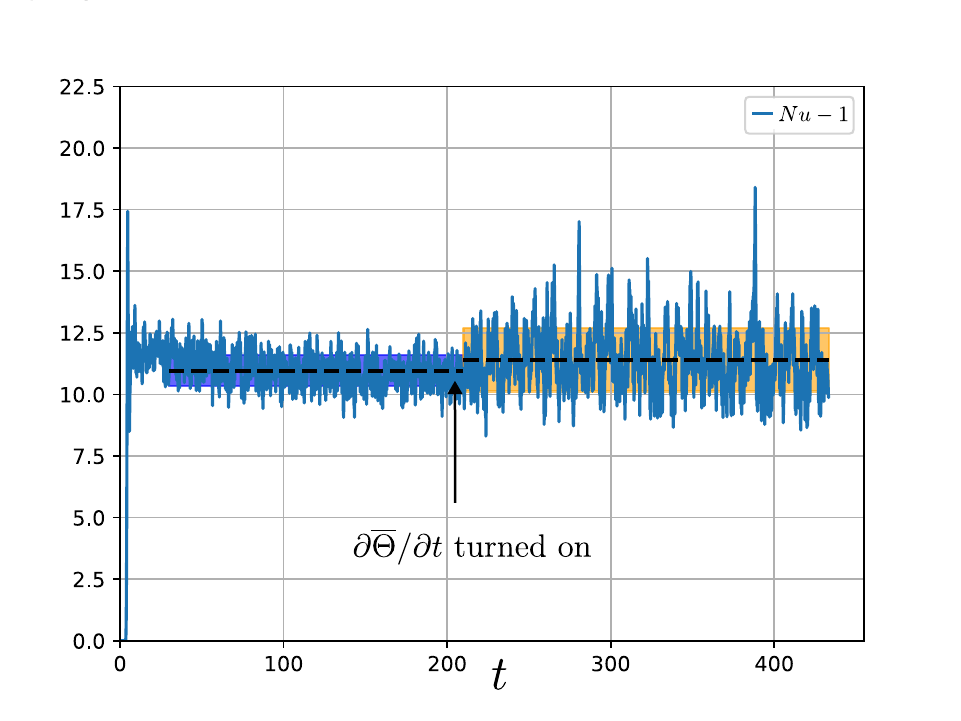}
   \includegraphics[width=0.32\textwidth]{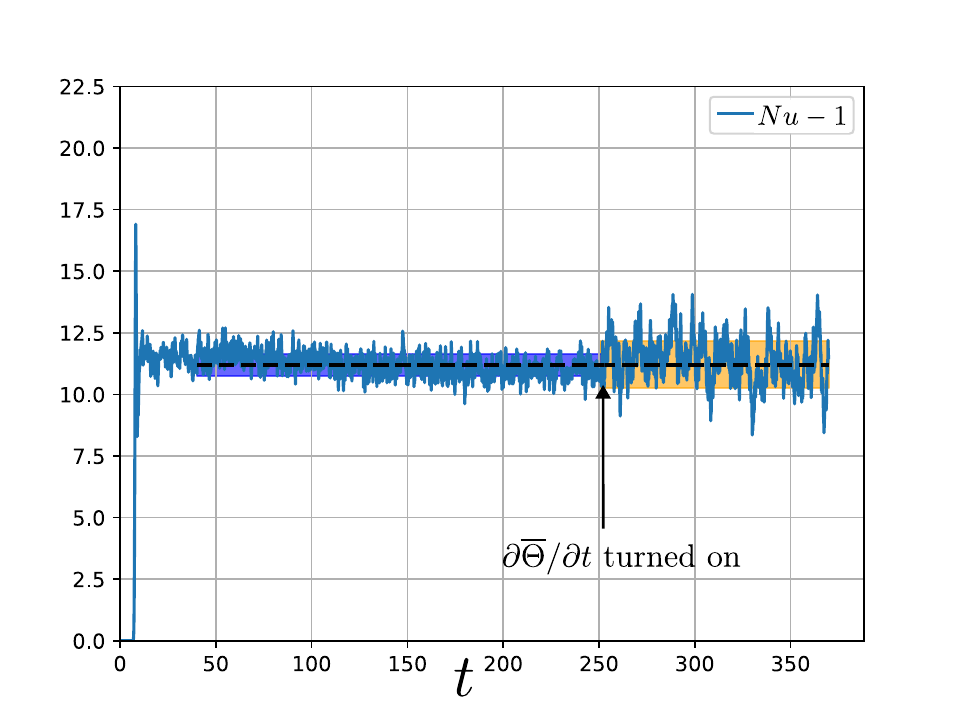}
   \includegraphics[width=0.32\textwidth]{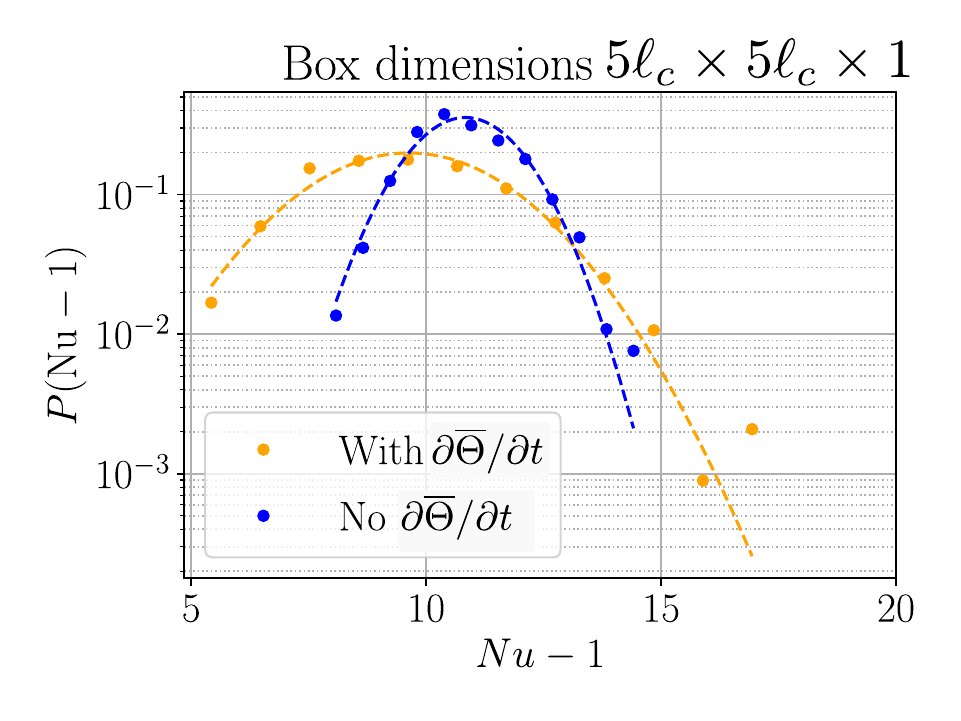}
   \includegraphics[width=0.32\textwidth]{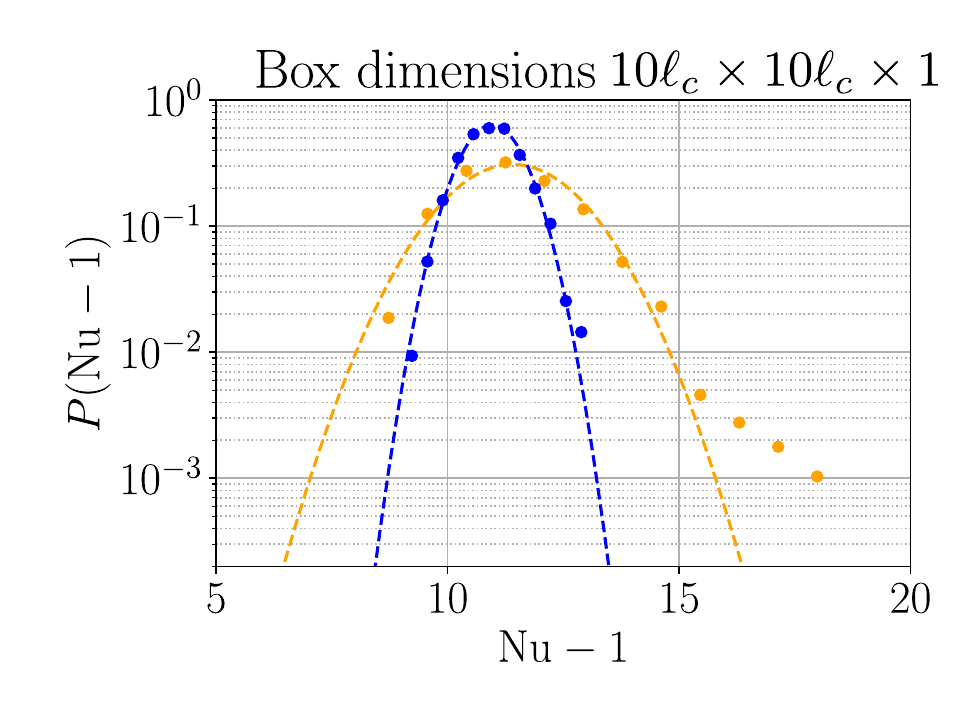}
   \includegraphics[width=0.32\textwidth]{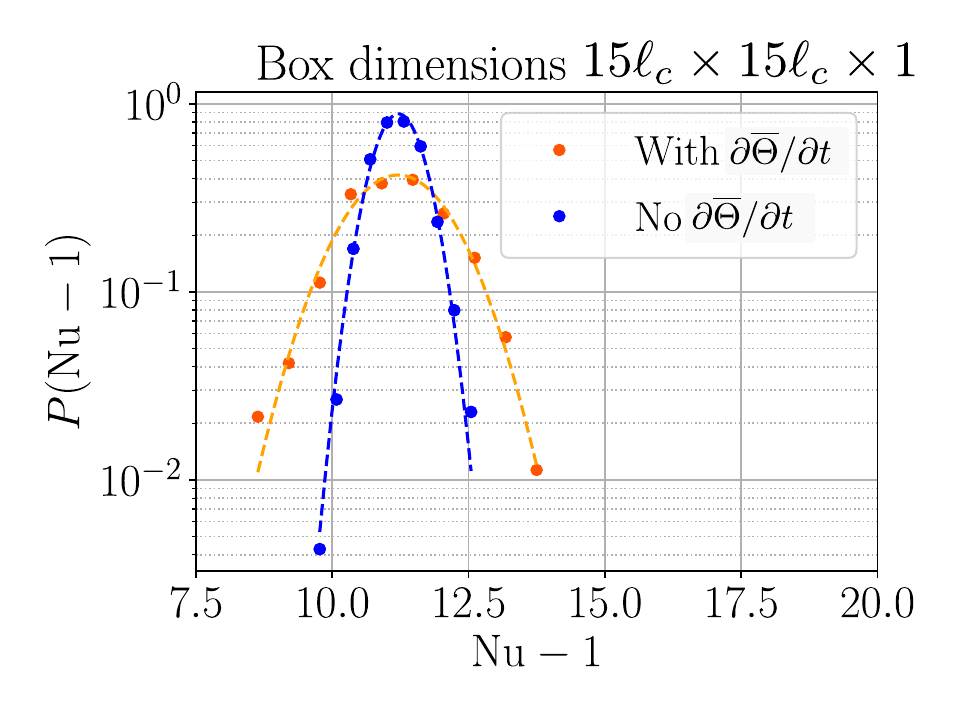}
  \caption{Panels (a)-(c): time series of $Nu-1$ at $\widetilde{Ra}=40$, $Pr=1$, $\ekman=10^{-9}$ for varying horizontal domain sizes. (a) $L=5\ell_c$ (resolution $64^2\times 128$ modes), (b) $L=10\ell_c$ (resolution $128^3$ modes), (c) $L=15\ell_c$ (resolution $192^3$ modes), with the rescaled most unstable length scale $\ell_c\approx 4.82$. Each simulation consists of two parts. In the first part, which extends from $t=0$ to the time $t=t_c$ indicated by a black arrow, the time derivative of $\overline{\Theta}$ in Eq.~(\ref{eq:rinse_Theta_mean}) is omitted, leading to a slaving relation between $\overline{\Theta}$ and the heat flux. At $t=t_c$ (at which the large scale vortex condensate reaches saturation), the time derivative of $\overline{\Theta}$ is restored. Horizontal dashed lines indicate time averages over each of the two segments of the simulation, while the blue and orange shaded areas indicate the observed standard deviation. The agreement between the results of the two numerical schemes improves noticeably as the domain size increases. Fluctuations are seen to be larger in smaller domains since the volume average contains fewer points. Panels (d)-(f): histograms of $Nu-1$ computed from each time series with and without the mean temperature time derivative, with dashed lines indicating a Gaussian fit. The histograms illustrate the convergence of the mean between the two schemes with increasing domain size. The histograms also show that fluctuations in the presence of the $\epsilon^{-2} \partial_t\overline{\Theta}$ term are of larger amplitude, a fact consistent with the time series in the top row.}
   \label{fig:dTmeandt_off and on}
\end{figure}
In the statistically steady state, quantities of interest, such as the Nusselt and Reynolds numbers are typically given as space-time averages. It is natural to expect that this averaging improves as the domain size is increased. This motivates the hypothesis, discussed in \ref{sec:tempomit}, that the term $\epsilon^{-2}\partial_t\overline{\Theta}$ in the equation may become subdominant in calculating, for instance, the Nusselt number, provided the domain size is sufficiently large. The strategy \kj{of omitting the temporal variation of the mean temperature}, which we will refer to as the \textit{slaving strategy}, has the significant advantage of accelerating the convergence to the steady state at small $\epsilon$ by orders of magnitude, due to the fact that $\epsilon^{-2} \partial_t = \partial_T$, where $T=\epsilon^2 t$, is a derivative with respect to a slow time variable. The slaving approach has already been used successfully for the reduced equations of Section~\ref{sec:asymp_guide} in a number of works, including \cite{mS06,kJ98a,kJ12b,aR14}. Here, we begin with a detailed verification of the slaving approach for the full Boussinesq system in the \avk{RRRiNSE} formulation. 

Figures~\ref{fig:dTmeandt_off and on}(a)-(c) show the time series of the Nusselt number for long simulations in domains of three different sizes, $5\ell_c\times5\ell_c\times1$, $10\ell_c\times10\ell_c\times1$, and $15\ell_c\times15\ell_c\times1$ at $\widetilde{Ra}=40$, $Pr=1$ and $Ek=10^{-9}$. With this set of parameters, an inverse cascade of energy is observed, which leads to the accumulation of energy at large scales and the formation of a large-scale vortex dipole (LSV), also observed in the nonhydrostatic 
quasi-geostrophic equations (\ref{eqns:reduced}), cf. \citep{kJ12b,sM21,aR14}. Each of the three simulation sets consists of two parts: first, each set is initialized with small-amplitude initial conditions and integrated for a long time with the slaving approach, until the LSV has saturated. Then, at the time indicated in Figs.~\ref{fig:dTmeandt_off and on}(a)-(c) by an arrow, the time derivative of the mean temperature is again included, restoring the full \avk{RRRiNSE} equations, and the run is continued. In small domains there is a notable discrepancy between the slaving strategy and the solution of the full \avk{RRRiNSE} equations, but this discrepancy decreases as the domain size increases (see dashed lines). This is accompanied by a decrease in the statistical fluctuations about the mean Nusselt number, as expected given the improved horizontal averaging in larger domains. The histograms in Figs.~\ref{fig:dTmeandt_off and on}(d)-(f) correspond to each of the two parts of the time series above and illustrate both of these trends: the averages of the two \avk{histograms} 
approach each other as the domain size is increased, and the variance decreases, being somewhat larger for the full equations than with the slaving strategy. Thus, for large enough domains (horizontal domain size $L\gtrsim 10\ell_c$ in terms of the critical length scale $\ell_c$), the slaving scheme yields approximately the same answers for mean quantities as the unaltered equations. On the other hand, differences remain in the fluctuations about that mean, owing to the additional slow time scale arising from the time derivative of the mean temperature, eliminated in the slaving strategy. We also note that in all cases, the peak of the histogram is close to a Gaussian, while in the presence of $\epsilon^{-2} \partial_t \overline{\Theta}$ deviations from that shape are seen, most strikingly in panel (e) in the intermediate domain, leading to a certain degree of skewness. This is not observed to the same degree in the smaller or larger domain, and remains to be better understood in future investigations.

\begin{figure}
   \centering
   \includegraphics[width=0.32\textwidth]{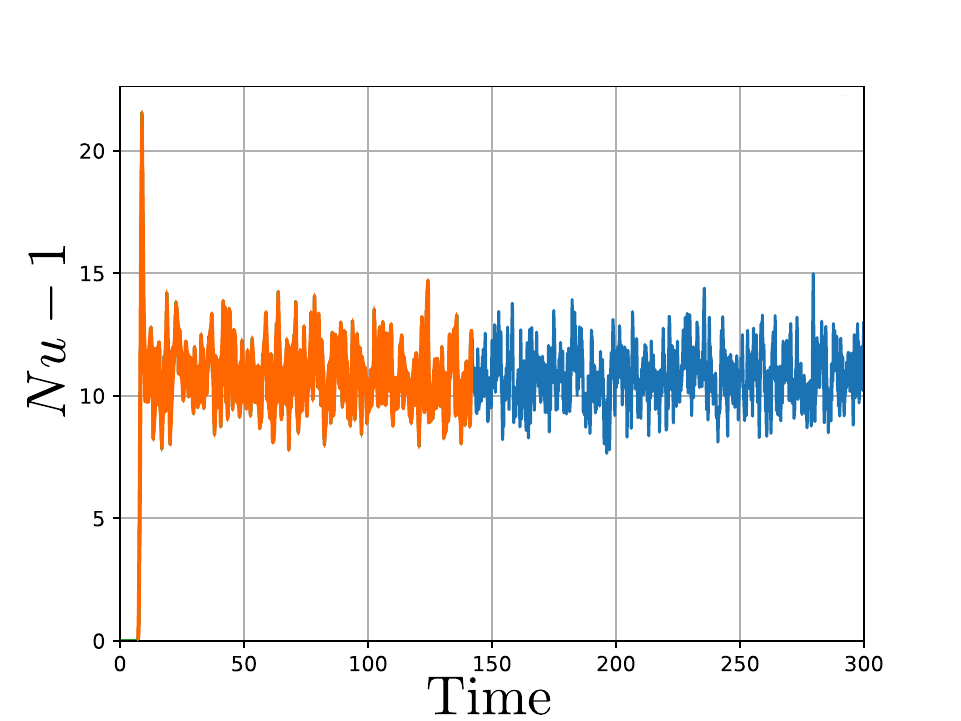}
   \includegraphics[width=0.32\textwidth]{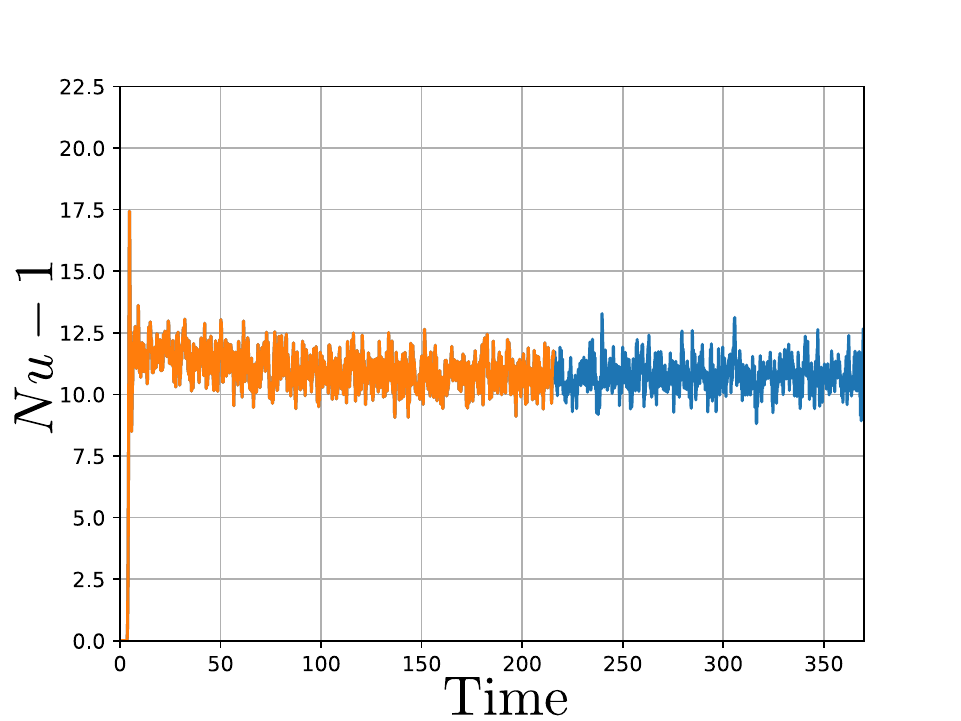}
   \includegraphics[width=0.31\textwidth]{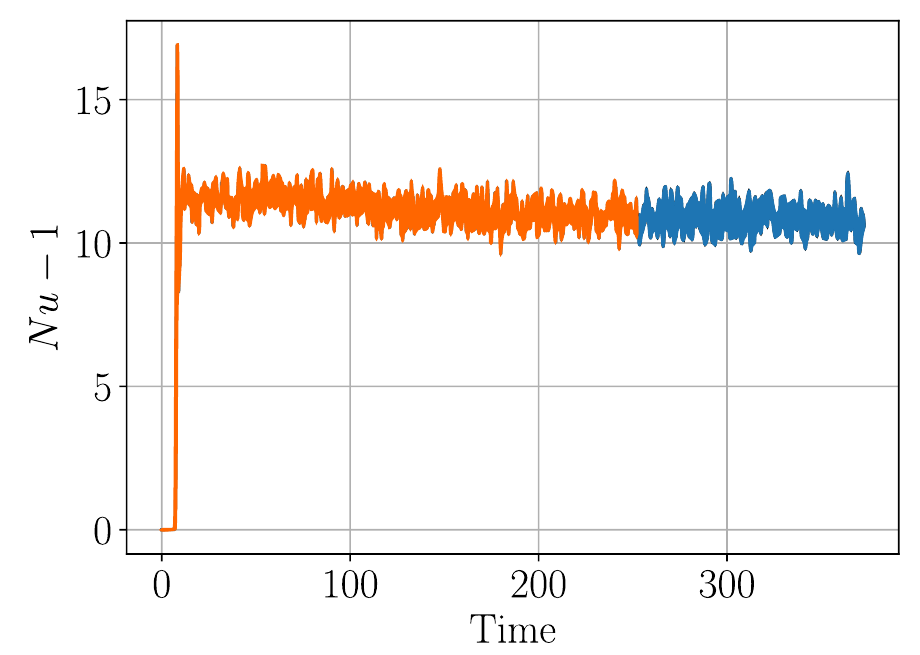}
  \caption{Time series of $Nu-1$ for various horizontal domain sizes at $Ek=10^{-9}$. (a) $L=5\ell_c$ (resolution $64^2\times 128$). (b) $L=10\ell_c$ (resolution $128^3$). (c) $L=15\ell_c$ (resolution $192^3$). First part of each simulation is performed without the $\partial_t\overline{\Theta}$ term in the mean temperature equation. In the second part of each run, the term $\epsilon^{-2}\partial_t\overline{\Theta}$ is replaced by $\partial_t\overline{\Theta}$. The mean and variance of each of the two segments are found to be very close. } \label{fig:time_series_dTmeandt_off_on_eps1}
\end{figure}

Figure~\ref{fig:time_series_dTmeandt_off_on_eps1} shows long time series similar to those in Figs.~\ref{fig:dTmeandt_off and on}(a)-(c), again at $Ek=10^{-9}$. The first, earlier part of each simulation (orange curves in Fig.~\ref{fig:time_series_dTmeandt_off_on_eps1}) is identical between the two figures and was computed using the slaving approach. In contrast, the second part (in blue) was computed with $\epsilon^{-2}\partial_t\overline{\Theta}$ replaced by $\partial_t\overline{\Theta}$ (without the $\epsilon^{-2}$ prefactor). The Nusselt number evolution in the two segments is found to be close to indistinguishable. This indicates that, in the statistically stationary state,  $\partial_t\overline{\Theta}$ is small compared to the remaining terms in the mean temperature equation.

In summary, the slaving strategy is a highly attractive scheme for accelerating transient dynamics in the approach to a statistically steady state, and it is often preferable to adopt this strategy for a sizeable efficiency gain. We therefore adopt the slaving strategy in all the runs described below. \avk{In \ref{app:comparison_slaving_rescaling}, we examine the impact of rescaling and slaving on the stability of strongly nonlinear simulations, showing that the slaving strategy yields a significant improvement in stability.}

\subsection{Implicit and explicit vertical derivatives in the diffusion terms}
\label{ssec:imp_exp_vert_deriv_diffusion}
A particular feature of the \avk{RRRiNSE} equations is the appearance of vertical derivatives with a prefactor of $\epsilon=\ekman^{1/3}$. This indicates that, for sufficiently small $\epsilon$, terms involving vertical derivatives become subdominant and their numerical treatment becomes {irrelevant in the \avk{Courant-Friedrichs-Lewy (CFL)} constraint} on the integration of the equations of motion. 

To test whether this intuition is correct, we perform two sets of runs at $\widetilde{Ra}=60$, $Pr=1$, varying $\ekman$ between $10^{-1}$ and $10^{-12}$, as listed in Table~\ref{tab:imp_exp_comp}. In the first set of runs, all diffusive terms are treated implicitly (as in all other runs described in later sections), while in the second set the diffusive terms in Eqs.~(\ref{eq:rinse_umom})-(\ref{eq:rinse_Theta_mean}) involving a vertical derivative are treated explicitly (the CFL condition is still only applied based on the horizontal velocity field). This is expected to produce no significant difference in the simulation outcome, provided $\ekman$ is sufficiently small.

\begin{figure}
\includegraphics[width=0.5\textwidth]{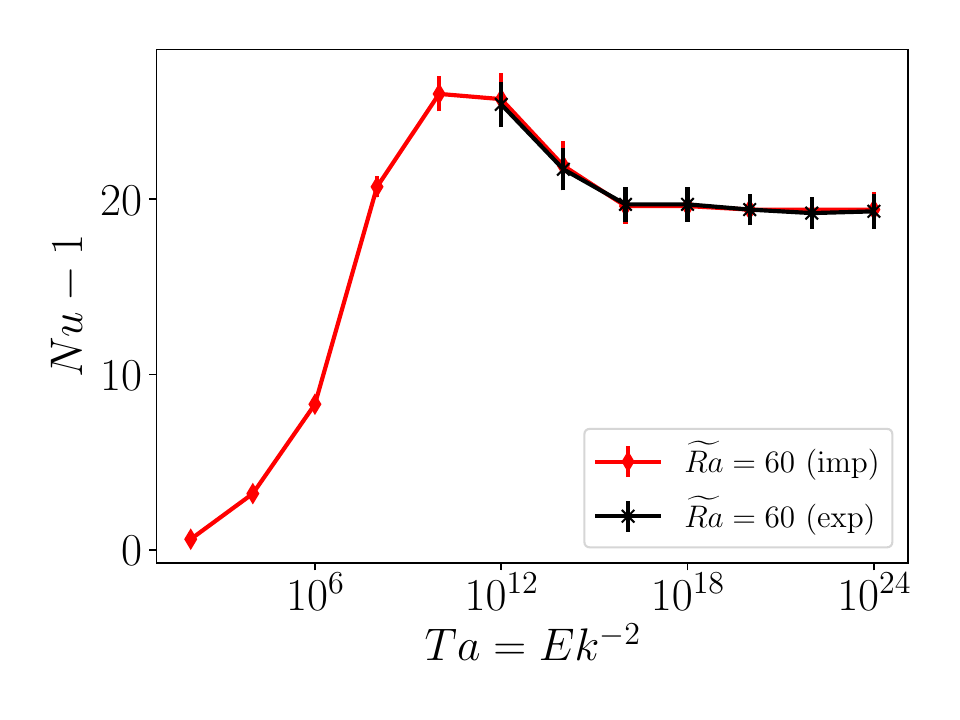}
\includegraphics[width=0.5\textwidth]{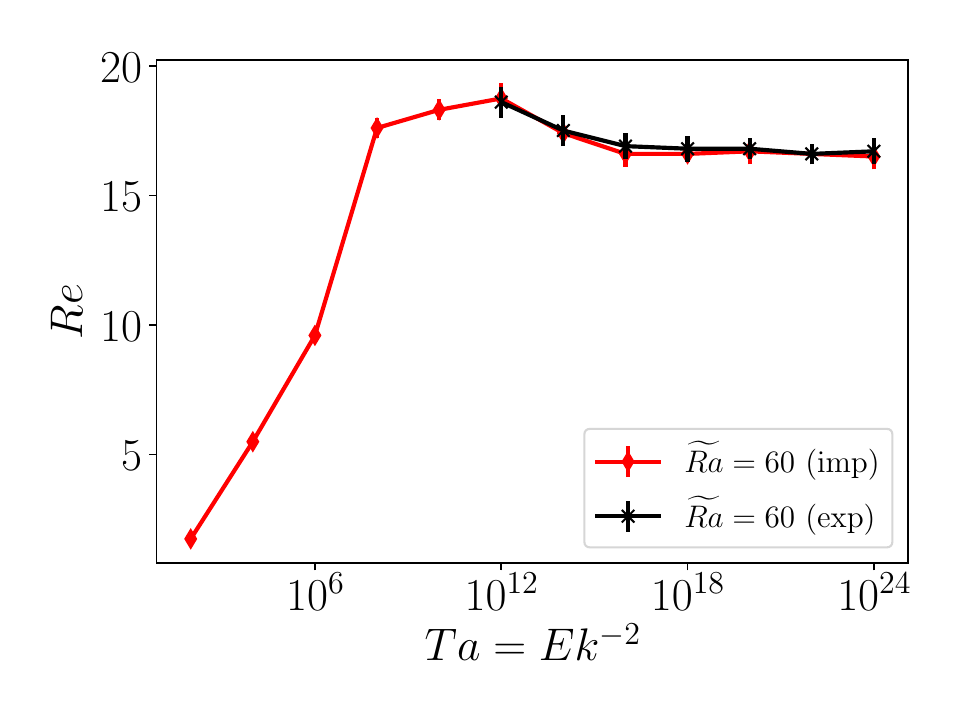}
\caption{Visualization of the data listed in Table \ref{tab:imp_exp_comp}, comparing the results of implicit and explicit vertical diffusion schemes at different Taylor numbers $Ta=Ek^{-2}$. At $Ta\geq 10^{12}$, implicit and explicit diffusion schemes are both stable  and yield very similar results which are compatible within the margin of error (given by the standard deviation). }
\label{fig:comp_imp_exp} 
\end{figure}

The data provided in Table~\ref{tab:imp_exp_comp} and its visualization in Fig.~\ref{fig:comp_imp_exp} show that the runs with explicit and implicit vertical diffusion schemes produce Nusselt and Reynolds numbers which are compatible with each other within the margin of error (computed as the standard deviation in steady state), provided that $\ekman=1/\sqrt{Ta}\lesssim 10^{-6}$. For $\ekman \gtrsim 10^{-6}$, i.e. $\epsilon \gtrsim 0.01$, the simulation becomes unstable with a CFL prefactor of $0.2$ and a time step tolerance of $0.3$, leading to an unphysical blow-up.

In short, diffusion terms involving vertical derivatives become irrelevant in the limit of small $\ekman$. This is consistent with the discussion in Section~\ref{sec:asymp_guide}, which highlighted that, as $\ekman\to0$, the \avk{RRRiNSE} formulation directly converges to the asymptotically reduced equations, which do not contain vertical diffusion terms except in the mean temperature equation.

\begin{table}
\small
\begin{center}
\begin{tabular}{|c|c|c|c|c|c|c|c|}
\hline
$\partial_Z$ (diffusion)  & $\ekman$  &  $\widetilde{Ra}$ & Ra &  $N_x \times N_y \times N_Z$ & Stability & $(Nu-1)\pm \Delta Nu$ & $Re_w\pm \Delta Re_w$\\ \hline
Implicit& $1.0\times10^{-1}$ & $60$ & $1.29\times10^3$  & $128 \times 128 \times 128$ & Stable &$0.6\pm0.1$ &$1.8\pm0.1$\\ \hline
Implicit & $1.0\times10^{-2}$ & $60$ & $2.78\times10^4$& $128 \times 128 \times 128$ & Stable &$3.2\pm0.2$ &$5.5\pm0.1$\\ \hline
Implicit & $1.0\times10^{-3}$& $60$ &$6.0\times10^5$ & $128 \times 128 \times 128$ & Stable &$8.3\pm0.3$ &$9.1\pm0.2$\\ \hline
Implicit  & $1.0\times10^{-4}$ & $60$& $1.29\times10^7$& $128 \times 128 \times 128$ & Stable & $20.7\pm0.6$&$17.7\pm0.3$\\ \hline
Implicit & $1.0\times10^{-5}$ & $60$  &$2.78\times10^8$& $128 \times 128 \times 128$ & Stable & $26.1\pm1.0 $&  $18.3\pm0.4$\\ \hline
Implicit & $1.0\times10^{-6}$ & $60$ & $6.00\times10^9$& $128 \times 128 \times 128$ & Stable & $25.4\pm 1.4$  & $18.8\pm0.7$\\ \hline
Implicit  & $1.0\times10^{-7}$& $60$ &$1.29\times10^{11}$ & $128 \times 128 \times 128$ & Stable & $21.9\pm1.4$ &$17.4\pm0.5$ \\ \hline
Implicit  & $1.0\times10^{-8}$& $60$& $2.78\times10^{12}$ & $128 \times 128 \times 128$ & Stable & $19.6\pm 1.0$ & $16.6\pm0.5$ \\ \hline
Implicit & $1.0\times10^{-9}$& $60$  & $6.00\times10^{13}$& $128 \times 128 \times 128$ & Stable & $19.6\pm0.9 $& $16.6\pm0.5$\\ \hline
Implicit  & $1.0\times10^{-10}$& $60$ &$1.29\times10^{15}$& $128 \times 128 \times 128$ & Stable &$19.4\pm1.0$ & $16.7\pm0.5$\\ \hline
Implicit& $1.0\times10^{-12}$ & $60$ &$6.00\times10^{17}$& $128 \times 128 \times 128$ & Stable & $19.5 \pm0.9$ & $16.5\pm0.5$\\ \hline \hline
Explicit & $1.0\times10^{-1}$& $60$ & $1.29\times10^3$ & $128 \times 128 \times 128$ & Unstable & -&- \\ \hline
Explicit & $1.0\times10^{-2}$ & $60$ & $2.78\times10^4$ & $128 \times 128 \times 128$ & Unstable & -&- \\ \hline
Explicit  & $1.0\times10^{-3}$ & $60$ & $6.00\times10^5$  & $128 \times 128 \times 128$ & Unstable & -&- \\ \hline
Explicit& $1.0\times10^{-4}$  & $60$ & $1.29\times10^7$ & $128 \times 128 \times 128$ & Unstable & - &- \\ \hline
Explicit  & $1.0\times10^{-5}$ & $60$ & $2.78\times10^8$  & $128 \times 128 \times 128$ & Unstable & - & - \\ \hline
Explicit  & $1.0\times10^{-6}$ & $60$ & $6.00\times10^9$ & $128 \times 128 \times 128$ & Stable & $25.4 \pm 1.3$ & $18.6\pm0.6$\\ \hline
Explicit & $1.0\times10^{-7}$ & $60$ & $1.29\times10^{11} $ & $128 \times 128 \times 128$ & Stable & $21.7\pm 1.2$ &  $17.5\pm 0.6$\\ \hline
Explicit  & $1.0\times10^{-8}$& $60$ & $2.78\times10^{12}$ & $128 \times 128 \times 128$ &  Stable & $19.7 \pm 1.0$ & $16.9 \pm 0.5$ \\ \hline
Explicit & $1.0\times10^{-9}$  & $60$ & $6.00\times10^{13}$ & $128 \times 128 \times 128$ & Stable &$19.6\pm1.0$& $16.8\pm0.5$ \\ \hline
Explicit  & $1.0\times10^{-10}$ & $60$ & $1.29\times10^{15}$ & $128 \times 128 \times 128$ & Stable &$19.4\pm0.9$& $16.8\pm0.5$ \\ \hline
Explicit & $1.0\times10^{-11}$ & $60$ & $2.78\times10^{16}$ & $128 \times 128 \times 128$ & Stable  &$19.2\pm0.9$& $16.6\pm0.4$ \\ \hline
Explicit  & $1.0\times10^{-12}$ & $60$ & $6.00\times10^{17}$& $128 \times 128 \times 128$ & Stable &$19.3\pm1.0$& $16.7\pm0.5$  \\ \hline
\end{tabular}
 \end{center}
\caption{Overview of runs with $\widetilde{Ra}=60$, $Pr=1$ and implicit or explicit vertical diffusion schemes (in a rescaled domain of size $10\ell_c \times 10\ell_c\times 1$, where $\ell_c = 4.82$) for Ekman numbers between $10^{-12}$ and $10^{-1}$. Values of $Nu-1$ and $Re_w$, defined in Eq.~(\ref{eq:def_Nu_Re}), refer to the average in the quasi-steady state during the early phase of the nonlinear evolution where a large-scale vortex condensate slowly grows in amplitude if an inverse cascade is present. Uncertainties represent the standard deviation of the time series.}
\label{tab:imp_exp_comp}
\end{table}

\subsection{Comparison with published Nusselt numbers}
\label{ssec:comparison_published_Nu}
\begin{figure}[h]
   \centering
   \includegraphics[width=0.6\textwidth]{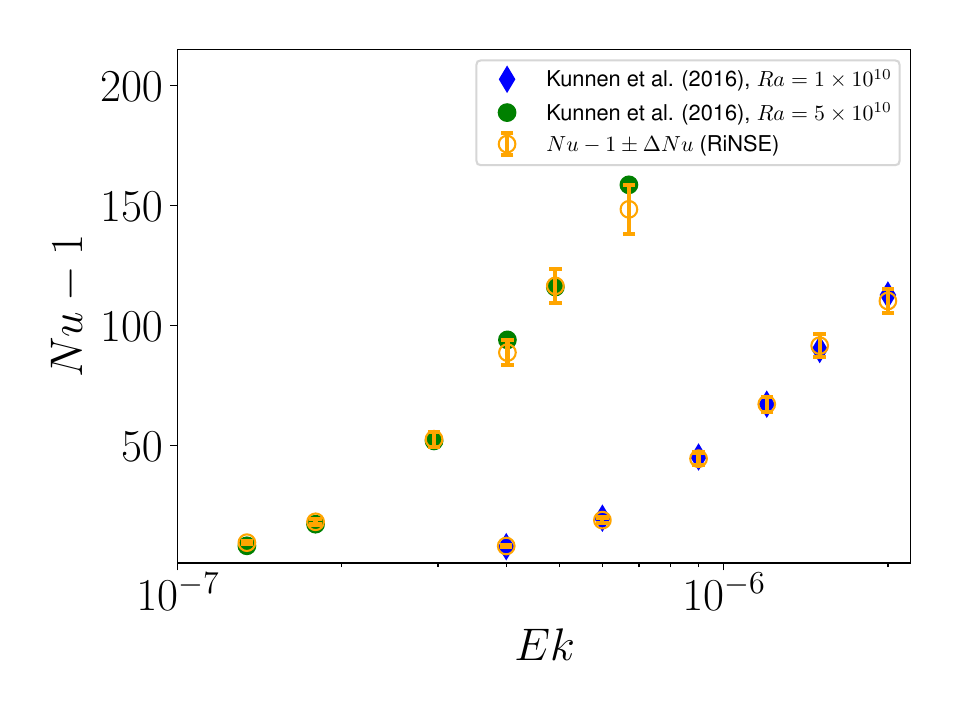}
   \caption{Comparison of Nusselt numbers from \cite{kunnen2016transition} with the \avk{RRRiNSE} results. The simulations shown correspond to runs A1--A12 in Table \ref{tab:parameters_runs_AB}. Error bars for \avk{RRRiNSE} data represent the observed standard deviation.}
   \label{fig:kunnen_comp}
\end{figure}

To further ascertain the validity of the \avk{RRRiNSE} formulation, we reproduce results from the literature. In \cite{kunnen2016transition}, Kunnen and co-workers provide Nusselt numbers obtained from direct numerical simulations  of the same set-up as ours, using a second-order energy-conserving, finite-difference code with fractional time-stepping. We stress that, while the authors of \cite{kunnen2016transition} discuss the \textit{transition to geostrophic turbulence}, they are only able to reach relatively modest Ekman numbers, $\ekman\gtrsim 1.34\times 10^{-7}$, which is large relative to the values of $\ekman$ that can be reached using \avk{RRRiNSE}, as shown in Sections~\ref{ssec:imp_exp_vert_deriv_diffusion} and \ref{ssec:convergence_geostrophic}. Nonetheless, the results presented in \cite{kunnen2016transition} provide a valuable benchmark as one bookend at moderate $\ekman$. 

We perform runs at the same parameters as those given in \cite{kunnen2016transition}. In Table~\ref{tab:parameters_runs_AB}, set A, we list those runs along with the Nusselt and Reynolds numbers obtained from our simulations. We choose up to $384\times 384$ dealiased Fourier modes in the horizontal directions and up to $256$ dealiased Chebyshev modes in the vertical derivation, while Kunnen et al. \cite{kunnen2016transition} consider up to $512 \times 512 \times 1024$ spatial grid points. We stress that these numbers cannot be easily compared between a spectral code such as \coral and finite-difference codes such as that employed in \cite{kunnen2016transition}. \textcolor{black}{However, the differences reside in the exponential vs algebraic error convergence properties of the two algorithmic approaches.}

Despite the different codes and resolution requirements, Fig.~\ref{fig:kunnen_comp} shows that the Nusselt numbers obtained using \avk{RRRiNSE} and those of Kunnen et al. agree well within the margin of error (the standard deviation of the Nusselt number time series). This provides a first bookend at relatively large Ekman numbers, where the \avk{RRRiNSE} formulation correctly reproduces known results.

\begin{figure}[h]
    \centering
 \includegraphics[width=0.6\textwidth]{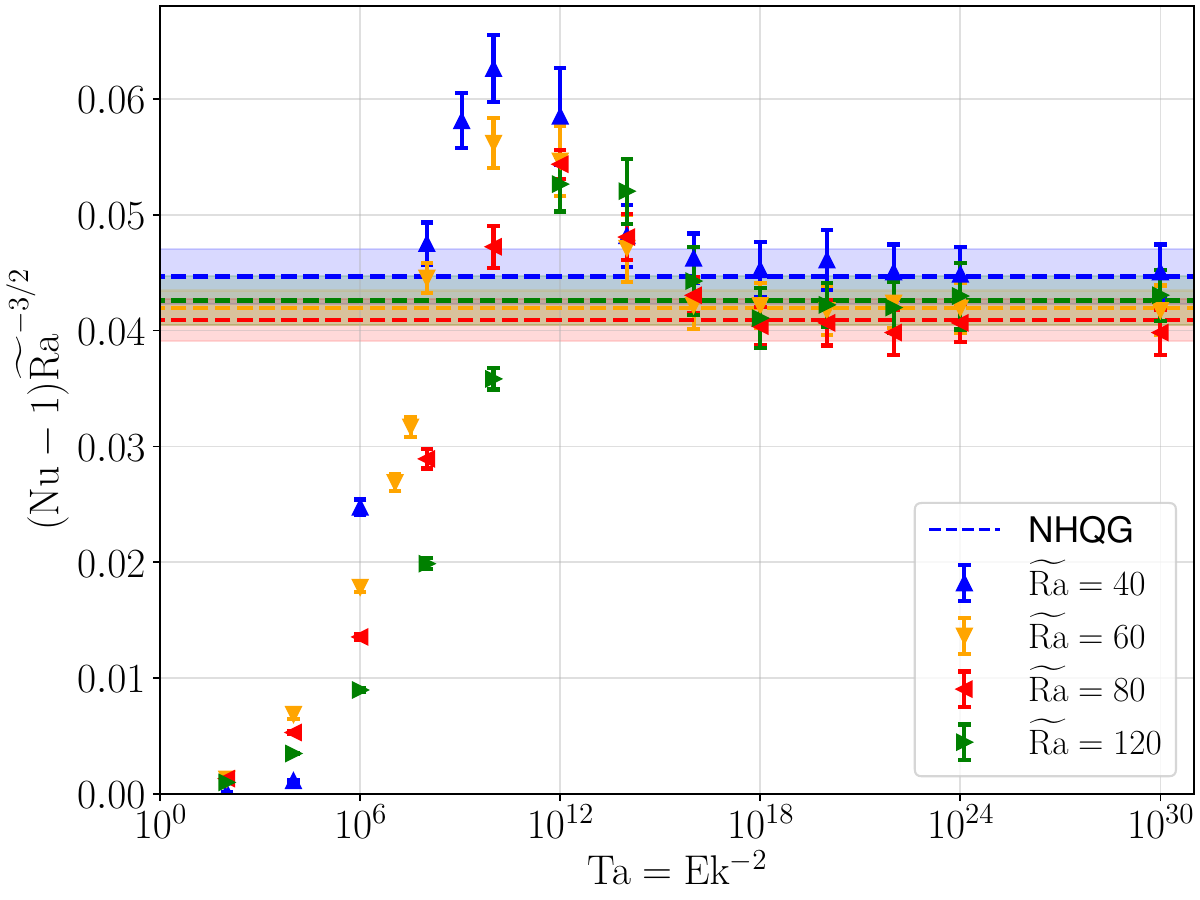}
    \caption{Nondimensional heat flux measured by the Nusselt number $Nu$, defined in Eq.~(\ref{eq:def_Nu_Re}), compensated by the turbulent scaling law $\widetilde{Ra}^{3/2}$, cf.~\cite{kJ12}, versus the Taylor number $Ta=\ekman^{-2}$ for fixed $\widetilde{Ra}\equiv Ra \ekman^{4/3}$ (\textcolor{black}{blue $\widetilde{Ra}=40$, orange $\widetilde{Ra}=60$, red $\widetilde{Ra}=80$, green $\widetilde{Ra}=120$}), computed with implicit vertical diffusion.  Dashed lines show the average in steady state predicted by the reduced equations (\ref{eqn:va})-(\ref{eqn:mta}) with the shaded region indicating one standard deviation about the mean. At low Ekman numbers, the \avk{RRRiNSE} predictions converge to the reduced equations.}
    \label{fig:Nu_vs_Ek}
\end{figure}
\subsection{Convergence to the geostrophic branch}
\label{ssec:convergence_geostrophic}
As discussed in Section~\ref{sec:asymp_guide}, it is expected that rotating convective flows converge to the well-studied, asymptotically reduced equations in the limit of small Ekman numbers. However, to date, it has not been possible to achieve sufficiently small Ekman numbers in direct numerical simulations of the full Boussinesq equations to observe this convergence.

Owing to the improved conditioning of the \avk{RRRiNSE} formulation in the small $\ekman$ limit, it is shown below that it becomes possible, for the first time, to observe this convergence. We perform four sets of simulations of the \avk{RRRiNSE} at $\widetilde{Ra}=40$, $60$, $80$, $120$, $Pr=1$ and $\ekman$ varying between $10^{-1}$ and $10^{-15}$ (corresponding to $Ta$ between $10^2$ and $10^{30}$), summarized in Tables \ref{tab:parameters_runs_AB}, \ref{tab:parameters_runs_CD} and \ref{tab:parameters_runs_EF}. We note that the \avk{RRRiNSE} formulation remained numerically stable even at $Ek=10^{-24}$, and yielded approximately the same Nusselt and Reynolds numbers as the case $Ek=10^{-15}$, but in order to avoid potential issues due to machine precision, these results are not shown here. In addition, we perform simulations of the asymptotically reduced equations described in Section~\ref{sec:asymp_guide} with $Pr=1$ and $\widetilde{Ra}=40$, $60$, $80$, $120$ and compare the observed Nusselt and Reynolds numbers with the \avk{RRRiNSE} results. 

\begin{figure}[h]
    \centering
      \includegraphics[width=0.6\textwidth]{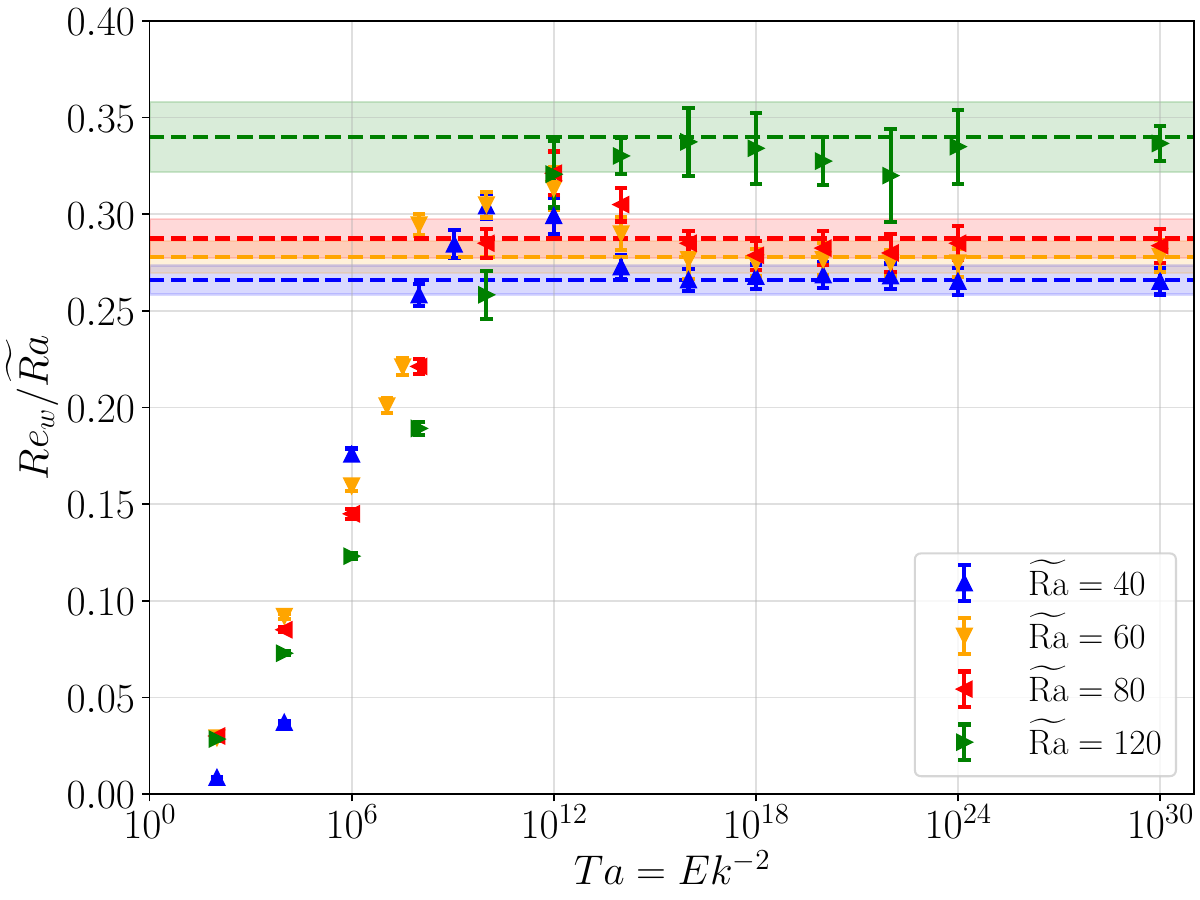}
    \caption{Reynolds number $Re_w$, defined in Eq.~(\ref{eq:def_Nu_Re}), 
    versus $Ta=\ekman^{-2}$ at fixed  $\widetilde{Ra}\equiv Ra \ekman^{4/3} = 60$. The dashed lines show the corresponding mean value obtained from the asymptotically reduced equations, with the shaded area indicating one standard deviation above and below that value.}
    \label{fig:Re_vs_Ek}
\end{figure}

Figures~\ref{fig:Nu_vs_Ek} and \ref{fig:Re_vs_Ek} show, respectively, the Nusselt and Reynolds numbers (based on vertical velocity and domain height), obtained at each $\widetilde{Ra}$ as a function of $Ta$ (symbols), with error bars indicating the observed standard deviation of the time series. The Nusselt number is rescaled by $\widetilde{Ra}^{3/2}$, \textcolor{black}{the turbulent scaling law \citep{jmA20},}  leading to an approximate data collapse between different $\widetilde{Ra}$ at large $Ta$ (small $\ekman$), in agreement with the asymptotically reduced equations \cite{kJ12b}. The Reynolds numbers are rescaled by $\widetilde{Ra}$, leading to a less satisfactory collapse, which is known to be related to the presence of the inverse energy cascade \cite{sM21,tO23}.

\begin{table}
\small
\begin{center}
\begin{tabular}{|c|c|c|c|c|c|c|}
\hline
Simulation nr. & \ekman & $\widetilde{Ra}$ & Ra &  $N_x \times N_y \times N_Z$ &   $(Nu-1) \pm \Delta Nu$ & $Re_w\pm \Delta Re_w$ \\
\hline
A1 & $4.0 \times 10^{-7}$ &  $29.5$ &$1\times10^{10}$   & $128\times 128\times 128$ & $8.1 \pm 0.5$  & $7.6\pm 0.2$ \\
A2 & $6.0 \times 10^{-7}$ & $50.6$  &$1\times10^{10}$   & $128\times 128\times 128$ & $18.9 \pm 1.1$  & $15.3\pm 0.5$\\
A3 & $9.0 \times 10^{-7}$ & $86.9$  &$1\times10^{10}$   & $128\times 128\times 128$ & $44.5 \pm 2.6$  & $29.2\pm 1.3$ \\ 
A4 & $1.2\times10^{-6}$   & $127.5$ &$1\times10^{10}$   & $256\times 256\times 256$& $67.2 \pm 3.1$  & $38.5\pm1.6$ \\
A5 & $1.5\times10^{-6}$   & $171.5$ & $1\times10^{10}$  & $256\times 256\times 256$& $91.6 \pm 4.7$ & $53.1\pm2.4$ \\
\textcolor{black}{A6} & $2.0\times 10^{-6}$  & $252.2$ & $1\times10^{10}$ & \avk{$384\times 384\times 256$} & \avk{$110.2 \pm 5$} &\avk{$59.6\pm 2.5$}\\
A7 & $1.34\times 10^{-7}$    & $34.3$  & $5\times 10^{10}$ & $128\times 128\times 128$ & $9.5 \pm 0.6$  &  $9.1\pm0.3$ \\
A8 & $1.79\times 10^{-7}$ & $50.4$  & $5\times 10^{10}$ & $128\times 128\times 128$ & $18.2 \pm 1.1$ & $14.8\pm0.4$\\
A9 & $2.95\times 10^{-7}$ & $98.3$  & $5\times 10^{10}$ & $128\times 128\times 128$ & $52.6 \pm 3.1$ & $33.0\pm1.6$\\
A10 & $4.02\times10^{-7}$ & $148.3$ & $5\times10^{10}$  &$256\times256\times256$ & $88.7 \pm 5.2$  & $47.9\pm1.5$ \\
A11 & $4.92\times 10^{-7}$ & $194.6$ & $5\times10^{10}$   & $256\times 256\times 256$ & $116.5 \pm 7.2$ & $61.4\pm3.2$ \\
\textcolor{black}{A12} & $6.71\times10^{-7}$ & $293.6$ & $5\times10^{10}$  &\avk{$384\times 384\times 256$} & \avk{$148.4\pm 10.2$} & \avk{$78.0\pm4.2$} \\ \hline
B1  & $1.0\times10^{-1}$  & 40      &  $8.62\times 10^{2}$  & $128\times 128\times 128$ & $0.029 \pm 0.007$ & $0.34\pm0.02$ \\
B2  & $1.0\times10^{-2}$  & 40      &  $1.86\times10^{4}$   & $128\times 128\times 128$ & $0.287 \pm0.014$  & $1.48\pm0.03$ \\
B3  & $1.0\times10^{-3}$  & 40      &  $4.0\times10^5$      & $128\times 128\times 128$ & $6.26\pm0.17$     &  $7.0\pm0.1$\\
B4  & $1.0\times10^{-4}$  & 40      &  $8.62\times10^{6}$  & $128\times 128\times 128$  & $12.0\pm0.4$      & $10.3\pm0.2$ \\
B5  & $3.0\times10^{-5}$  & 40      &  $4.29\times10^{7}$  & $128\times 128\times 128$  & $14.7\pm0.4$      & $11.4\pm0.3$ \\
B6  & $1.0\times10^{-5}$  & 40      &  $1.86\times10^{8}$  & $128\times 128\times 128$  & $15.8\pm0.7$      & $12.2\pm0.3$ \\
B7  & $1.0\times10^{-6}$  & 40      &  $4.0\times10^{9}$   & $128\times 128\times 128$  & $14.8\pm1.0$      & $12.0\pm 0.4$  \\
B8  & $1.0\times10^{-7}$  & 40      &  $8.62\times10^{10}$ & $128\times 128\times 128$  & $12.2\pm0.7$      & $10.9\pm0.3$  \\
B9  & $1.0\times10^{-8}$  & 40      &  $1.86\times10^{12}$ & $128\times 128\times 128$  & $11.7\pm 0.5$     & $10.6\pm0.2$ \\
B10 & $1.0\times10^{-9}$  & 40      &  $4.0\times10^{13}$  & $128\times 128\times 128$  & $11.5\pm 0.6$     & $10.7\pm0.3$ \\
B11 & $1.0\times10^{-10}$ & 40      &  $8.62\times10^{14}$ & $128\times 128\times 128$  & $11.65\pm0.65$  & $10.7\pm0.3$\\
B12 & $1.0\times10^{-11}$ & 40      &  $1.86\times10^{16}$ & $128\times 128\times 128$  & $11.4\pm0.6$      & $10.7\pm0.3$ \\
B13 & $1.0\times10^{-12}$ & 40      &  $4.0\times10^{17}$  & $128\times 128\times 128$  & $11.4\pm0.6$      & $10.6\pm0.3$ \\
B14 & $1.0\times10^{-15}$ & 40      &  $4.0\times10^{21}$   & $128\times 128\times 128$  & $11.4\pm0.6$      & $10.6\pm0.3$\\
B15 & $1.0\times10^{-15}$ & 40      &  $4.0\times10^{21}$   & $256\times 256\times 256$  & $11.4\pm 0.5$     & $10.7\pm0.3$\\
\hline\end{tabular}
\end{center}
\caption{List of simulations described in this work (see also Tables \ref{tab:parameters_runs_CD} and \ref{tab:parameters_runs_EF}). All simulations are done with $Pr=1$ in a rescaled domain of size $10\ell_c \times 10\ell_c\times 1$, where $\ell_c \approx 4.82$. Simulations A1 through A12 have $Ra$, $\ekman$ and $\Pr$ identical to those in \cite{kunnen2016transition}. The resolution is specified by the numbers $N_x$, $N_y$ of Fourier modes in the horizontal directions and the number $N_Z$ of Chebyshev modes in the vertical. The values of $Nu$ and $Re_w$, defined in Eq.~(\ref{eq:def_Nu_Re}), refer to the average computed in the early, quasi-steady, nonlinear stage of the evolution, in the absence of an LSV. Uncertainties represent the standard deviation of the time series. }
\label{tab:parameters_runs_AB}
\end{table}%
\begin{table}
\small
\begin{center}
\begin{tabular}{|c|c|c|c|c|c|c|}
\hline
Simulation nr. & $\ekman$ & $\widetilde{Ra}$ & Ra &  $N_x \times N_y \times N_Z$ &   $(Nu-1) \pm \Delta Nu$ & $Re_w\pm \Delta Re_w$ \\
\hline
C1 &  $1.0\times10^{-1}$ & 60  & $1.29\times10^{3}$  & $128\times 128\times 128$ & $0.6\pm0.1$  & $ 1.8 \pm0.1$\\
C2 & $1.0\times10^{-2}$  & 60  &  $2.78\times10^{4}$  & $128\times 128\times 128$ & $3.2\pm0.2$  & $5.5 \pm 0.1$\\
C3 & $1.0\times10^{-3}$  & 60  &   $6.0\times10^{5}$ & $128\times 128\times 128$ & $8.3\pm0.3$  & $9.1\pm0.2$\\ 
C4 & $3.0\times10^{-4}$  & 60  &  $2.99\times10^{6}$  & $128\times 128\times 128$& $12.5 \pm 0.4$  & $12.1 \pm 0.2$\\
C5 & $1.75\times10^{-4}$  & 60  & $6.13\times10^{6}$  & $128\times 128\times 128$& $14.7 \pm 0.4$ & $13.3 \pm0.3$ \\
C6 & $1.0\times10^{-4}$  & 60  &  $1.29\times10^7$ & $128\times 128\times 128$ & $20.7 \pm0.6$ & $ 17.7\pm0.3$\\
C7 & $1.0\times10^{-5}$  & 60  & $2.78\times10^{8}$ & $128\times 128\times 128$ & $26.1 \pm 1.0$  & $18.3\pm0.4$ \\
C8 & $1.0\times10^{-6}$  & 60  & $6.0\times10^{9}$ & $128\times 128\times 128$ & $25.4 \pm 1.4$ & $18.8\pm0.7$\\
C9 & $1.0\times10^{-7}$  & 60  & $1.29\times10^{11}$ & $128\times 128\times 128$ & $21.9 \pm 1.4$ & $17.4\pm0.5$ \\
C10 & $1.0\times10^{-8}$ & 60  & $2.78\times10^{12}$ &$128\times 128\times 128$ & $19.6\pm 1.0$ & $16.6\pm0.6$\\
C11 & $1.0\times10^{-9}$ & 60  & $6.00\times10^{13}$   & $128\times 128\times 128$ & $19.6 \pm 0.9$  &$16.5\pm0.4 $\\
C12 & $1.0\times10^{-10}$  & 60  & $1.29\times10^{15}$ &$128\times 128\times 128$ & $19.4 \pm 1.0$& $16.6\pm0.5$\\
C13 & $1.0\times10^{-11}$ & 60 & $2.78\times10^{16}$ & $128\times128\times128$&  $19.7\pm 1.0$  & $16.5\pm0.4$\\
C14 & $1.0\times10^{-12}$  & 60  & $6.00\times10^{17}$ &$128\times 128\times 128$ & $ 19.5\pm 1.0$  & $16.5\pm0.4$\\
C15 & $1.0\times10^{-15}$  & 60  &$6.00\times10^{21}$  &$128\times 128\times 128$ & $19.4 \pm 1.0$ & $16.7\pm0.5$\\
\hline
D1  & $1.0\times10^{-1}$  & 80   & $1.72\times10^3$  & $192\times 192\times 192$  & $0.95\pm 0.05$   &  $2.4\pm0.1$\\
D2  & $1.0\times10^{-2}$  & 80   & $3.71\times10^4$  & $192\times 192\times 192$  & $3.8\pm0.1$      &  $6.8\pm0.1$ \\
D3  & $1.0\times10^{-3}$  & 80   & $8.00\times10^5$  & $192\times 192\times 192$  & $9.7\pm0.2$      &  $11.6\pm0.2$\\
D4  & $1.0\times10^{-4}$  & 80   & $1.72\times10^7$  & $192\times 192\times 192$  & $20.7\pm0.6$     &  $17.7\pm0.3$\\
D5  & $1.0\times10^{-5}$  & 80   & $3.71\times10^8$  & $192\times 192\times 192$  & $33.8\pm0.9$     &  $25.7\pm0.9$\\
D6  & $1.0\times10^{-6}$  & 80   & $8.00\times10^9$  & $192\times 192\times 192$  & $38.9\pm0.9$     &  $24.4\pm0.7$\\
D7  & $1.0\times10^{-7}$  & 80   & $1.72\times10^{11}$ & $192\times 192\times 192$  & $34.4\pm1.4$   &  $24.4\pm0.7$\\
D8  & $1.0\times10^{-8}$  & 80   & $3.71\times10^{12}$ & $192\times 192\times 192$  & $30.8\pm1.1$   &  $22.8\pm0.8$\\
D9  & $1.0\times10^{-9}$  & 80   & $8.00\times10^{13}$ & $192\times 192\times 192$  & $28.9\pm1.2$   &  $22.3\pm0.6$\\
D10 & $1.0\times10^{-10}$ & 80   & $1.72\times10^{15}$ & $192\times 192\times 192$  & $29.1\pm1.6$    &  $22.6\pm0.7$\\
D11 & $1.0\times10^{-11}$ & 80   & $3.71\times10^{16}$ & $192\times 192\times 192$  & $28.5\pm1.4$    &  $22.4\pm0.8$\\
D12 & $1.0\times10^{-12}$ & 80   & $8.00\times10^{17}$ & $192\times 192\times 192$  & $29.1\pm1.2$   &  $22.8\pm0.7$\\
D13 & $1.0\times10^{-15}$ & 80   & $8.00\times10^{21}$ & $192\times 192\times 192$  & $28.5\pm1.4$    &  $22.7\pm0.7$\\
\hline\end{tabular}
\end{center}
\caption{List of simulations described in this work (continued 1); see also Tables \ref{tab:parameters_runs_AB} and \ref{tab:parameters_runs_EF}. Simulations C1-C3, C6-13 are identical to the runs with an implicit vertical diffusion scheme listed in Table \ref{tab:imp_exp_comp}.}
\label{tab:parameters_runs_CD}
\end{table}%
\begin{table}
\small
\begin{center}
\begin{tabular}{|c|c|c|c|c|c|c|}
\hline
Simulation nr. & $\ekman$ & $\widetilde{Ra}$ & Ra &  $N_x \times N_y \times N_Z$ &   $(Nu-1) \pm \Delta Nu$ & $Re_w\pm \Delta Re_w$ \\
\hline
E1 & $1.0\times10^{-1}$ & 120&$2.59\times10^3$&$128\times128\times128$ & $1.3\pm0.1$& $3.4\pm0.1$\\
E2 & $1.0\times10^{-2}$ & 120&$5.57\times10^4$&$128\times128\times128$ & $4.6\pm0.1$& $8.7\pm0.1$ \\
E3 & $1.0\times10^{-3}$ & 120&$1.2\times10^6$&$128\times128\times128$& $11.8\pm0.2$& $14.8\pm0.2$\\
E4 & $1.0\times10^{-4}$ & 120&$2.59\times10^7$&$128\times128\times128$& $26.1\pm0.6$& $22.7\pm0.4$\\
E5 & $1.0\times10^{-5}$& 120&$5.57\times10^8$&$256\times256\times256$& $45.0\pm1.9$&$30.2\pm0.9$\\
E6 & $1.0\times10^{-6}$& 120& $1.2\times10^{10}$&$256\times256\times256$& $63.2\pm3.5$&$37.9\pm1.6$\\
E7 &$ 1.0\times10^{-7}$ & 120& $2.59\times10^{11}$&$256\times256\times256$ & $64.4\pm3.7$&$39.6\pm1.1$\\
E8 &$ 1.0\times10^{-8}$ & 120& $5.57\times10^{12}$&$256\times256\times256$ & $58.2\pm3.9$&$40.5\pm2.1$\\
E9 &$ 1.0\times10^{-9}$ & 120& $1.2\times10^{14}$&$256\times256\times256$& $54.0\pm3.4$&$40.1\pm2.2$\\
E10 & $1.0\times10^{-10}$  & 120 & $2.59\times10^{15}$&$256\times256\times256$& $54.9\pm2.5$&$39.7\pm1.5$\\
E11 & $1.0\times10^{-11}$  & 120 & $2.59\times10^{15}$&$256\times256\times256$ & $55.2\pm2.9$&$38.7\pm1.0$\\
E12 & $1.0\times10^{-12}$ & 120 & $1.20\times10^{18}$&$256\times256\times256$ & $56.5\pm3.8$&$40.3\pm1.1$\\
E13 & $1.0\times10^{-15}$  & 120 & $1.20\times10^{22}$&$256\times256\times256$ & $56.6\pm2.9$&$40.4\pm 1.0$\\
\hline
F1 & $\ekman\ll1$ & 40 & $Ra\gg1$ & $128\times128\times128$ & $11.3\pm1.0$ &$10.6\pm0.3$\\ 
F2 & $\ekman\ll1$& 60 & $Ra\gg1$ & $128\times128\times128$ & $19.5\pm0.7$ &$16.7\pm0.5$\\ 
F3 & $\ekman\ll1$ & 80 & $Ra\gg1$ & $128\times128\times128$ & $29.3\pm1.3$
& $23.1\pm0.8$
\\ 
F4 & $\ekman\ll1$ & 120 & $Ra\gg1$ & $256\times256\times256$ & $56.0\pm2.8$ &$40.9\pm2.1$\\
\hline
\end{tabular}
\end{center}
\caption{List of simulations described in this work (continued 2); see also Tables \ref{tab:parameters_runs_AB} and  \ref{tab:parameters_runs_CD}.}
\label{tab:parameters_runs_EF}
\end{table}

In addition, the dashed lines in Figs.~\ref{fig:Nu_vs_Ek} and \ref{fig:Re_vs_Ek} indicate the results obtained using the asymptotically reduced equations, with the shaded area showing the standard deviation. The \avk{RRRiNSE} results are seen to converge to the values observed in the asymptotically reduced equations above a certain threshold in the Taylor number $Ta$, within the error margins given by the standard deviation of the time series from the reduced equations. The threshold $Ta$ required for this convergence appears to increase with $\widetilde{Ra}$, but a more detailed investigation will be required in the future to quantitatively investigate this behavior. The observed convergence of the \avk{RRRiNSE} to the reduced equations provides an additional bookend validating the accuracy of the \avk{RRRiNSE} formulation against an established body of work in the limit $\ekman\to 0$ ($Ta\to\infty$). We also verified that the flow statistics obtained in the low $\ekman$ regime are independent of the time step using runs with CFL prefactor $0.4$ or $0.1$ instead of $0.2$ (which was used in all other runs) using $Ek=10^{-15}$, $\rRa = 80$, both of which gave the same Nusselt and Reynolds numbers within one standard deviation (not shown).

The above simulations were performed with between $128$ and $384$ dealiased Fourier modes in the $x$ and $y$ directions and between $128$ and $256$ dealiased Chebyshev modes in the vertical direction. It was verified for each simulation that the thermal boundary layer (defined in terms of the root-mean-square temperature fluctuation, cf. \cite{kJ12}) was well resolved, with at least 10 grid points. 

Beyond the convergence of the Nusselt and Reynolds numbers to the values predicted by the asymptotic equations, an interesting pattern emerges. Both $Nu$ and $Re$ are small when $Ta$ is small (weak rotation). As $Ta$ increases, $Nu$ and $Re$ increase as well and exhibit an overshoot before converging to the asymptotic value. The amplitude of the overshoot is seen to decrease with increasing $\widetilde{Ra}$. Similar results were very recently reported in \cite{anas2024critical} for high $\Pr$ rotating Rayleigh-B\'enard convection with no-slip boundary conditions, but the physical origin of these features remains to be explained. An enhancement of the Nusselt number with increasing rotation rate has also been observed for $Pr=4.38$ and $6.4$ \cite{hartmann2023optimal}. The \avk{RRRiNSE} formulation allows us, for the first time, to observe the full range of $\ekman$ from order one values down to the asymptotic regime within a single code, opening the door to detailed numerical studies of the classical problem of rapidly rotating convection, which has long posed a major challenge to the fluid dynamics community.

\subsection{Visualizations}
\label{ssec:vis}
\begin{figure}
    \centering
      \hspace{-1cm} Side view pressure $p$ \hspace{5.0cm} Side view $v$\\
    \includegraphics[width=0.49\textwidth]{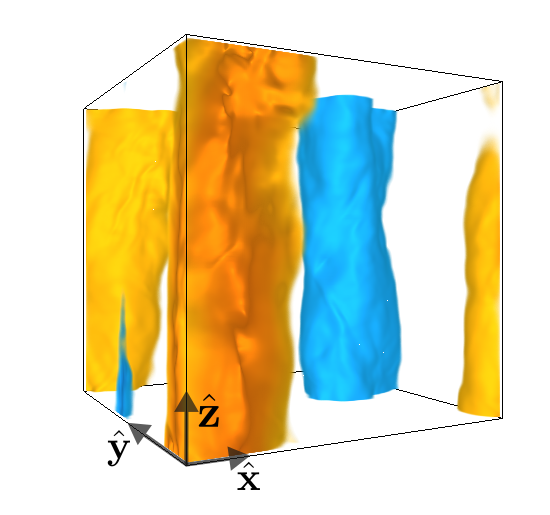}
    \includegraphics[width=0.49\textwidth]{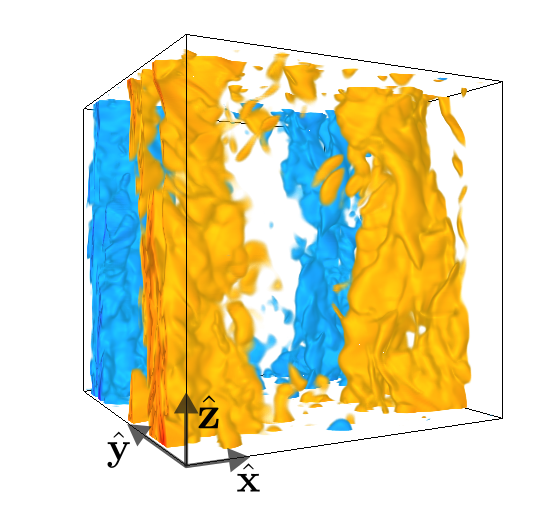}
   \\    \hspace{-1cm} Top view pressure $p$ \hspace{5.0cm} Top view $v$ \\
    \includegraphics[width=0.49\textwidth,height=0.45\textwidth]{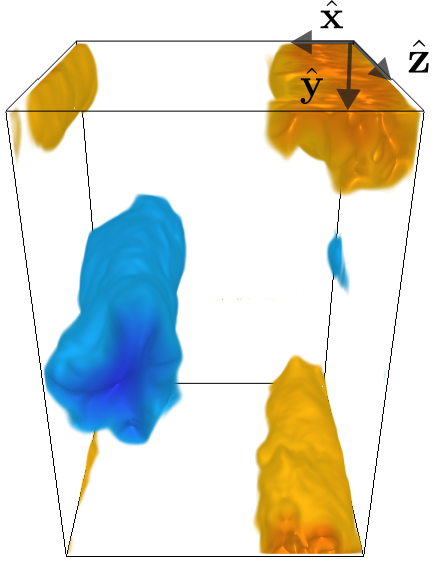}
    \includegraphics[width=0.49\textwidth,height=0.45\textwidth]{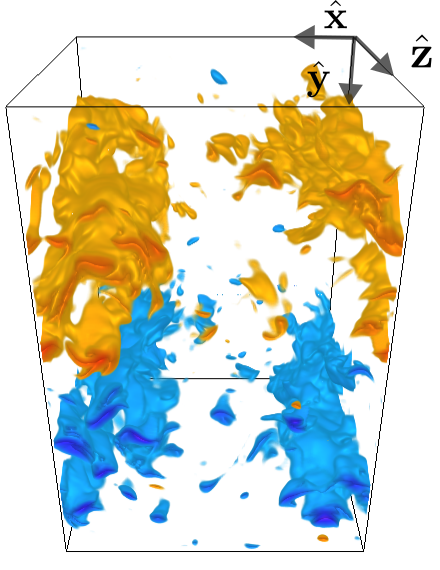}
    \caption{Snapshots of the pressure (left column) and $y$-component $v$ of the velocity (right column) from run D12 (with $\ekman=10^{-15}$, $\widetilde{Ra}=80$, $Pr=1$) in the steady state, where a saturated LSV is present. The axes in all panels are indicated by black arrows. Top row: side view. Bottom row: same data as in the top row (viewed from top). Blue color indicates negative values while orange and red colors indicate positive values.}
\label{fig:vis_press_omz}
\end{figure}
\begin{figure}
    \centering
          \hspace{-1cm} Side view $\omega_z$ \hspace{5.0cm} Side view $w$\\
     \includegraphics[width=0.49\textwidth]{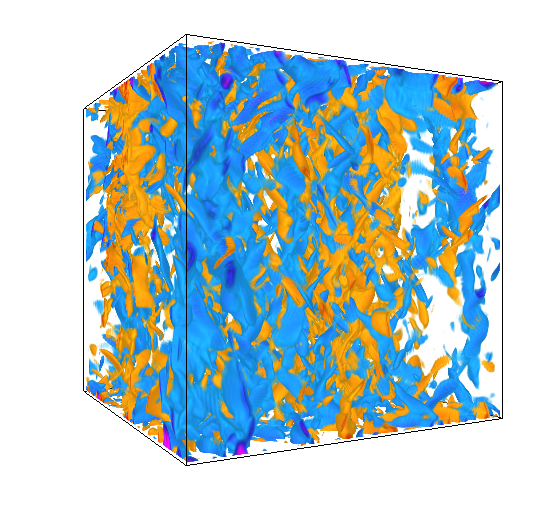}
    \includegraphics[width=0.49\textwidth]{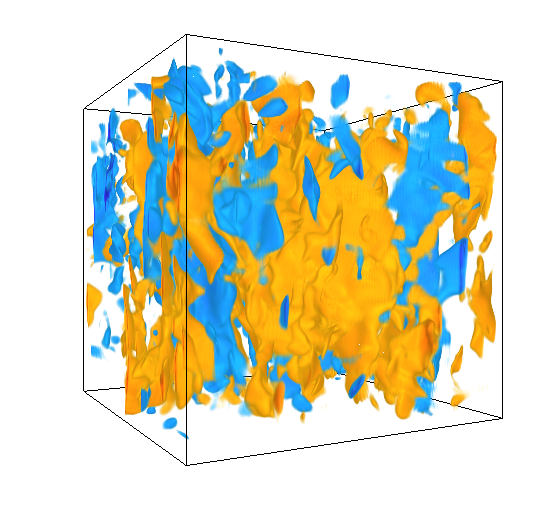}\\
              \hspace{-1cm} Top view $\omega_z$ \hspace{5.0cm} Top view $w$\\
        \includegraphics[width=0.49\textwidth,height=0.47\textwidth]{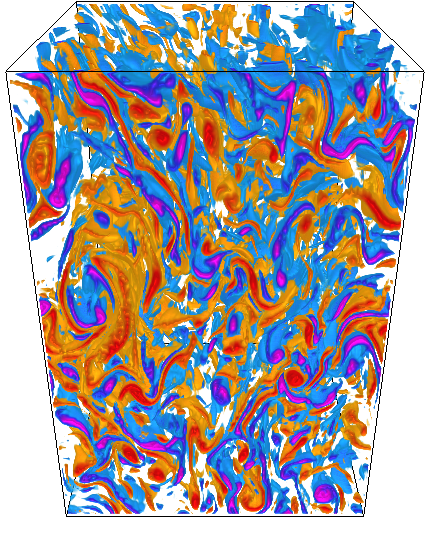}
    \includegraphics[width=0.49\textwidth,height=0.47\textwidth]{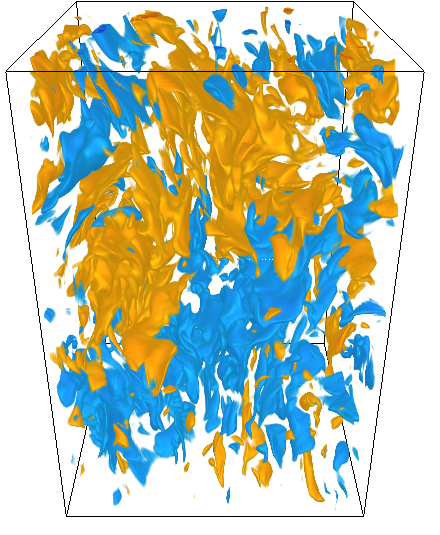}\\
    \caption{Snapshots of the vertical vorticity $\omega_z$ (left column) and the vertical velocity $w$ from run D12 (with $\ekman=10^{-15}$, $\widetilde{Ra}=80$, $Pr=1$) in the steady state, where a saturated LSV is present. The orientation in all panels is identical to Fig. \ref{fig:vis_press_omz}. Top row: side view of $\omega_z$ (left) and $w$ (right). Bottom row: same data as in top row (viewed from top). Blue color indicates negative values while orange and red colors indicate positive values.  }
\label{fig:vis_velx_velz}
\end{figure}
\begin{figure}
    \centering
    \hspace{-1cm} Side view $\theta$ \hspace{5.0cm} Side view $U$\\
    \includegraphics[width=0.49\textwidth]{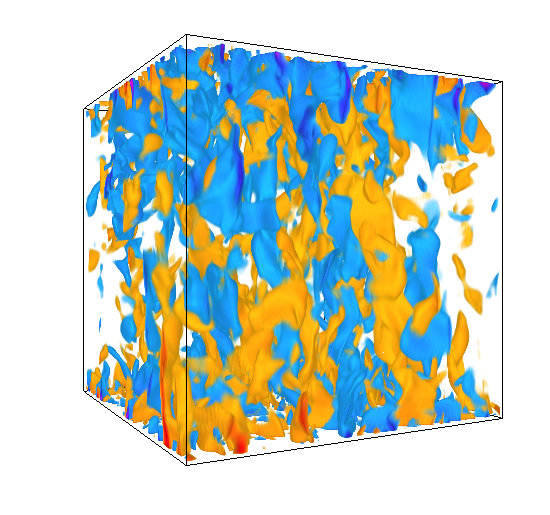}
    \includegraphics[width=0.49\textwidth]{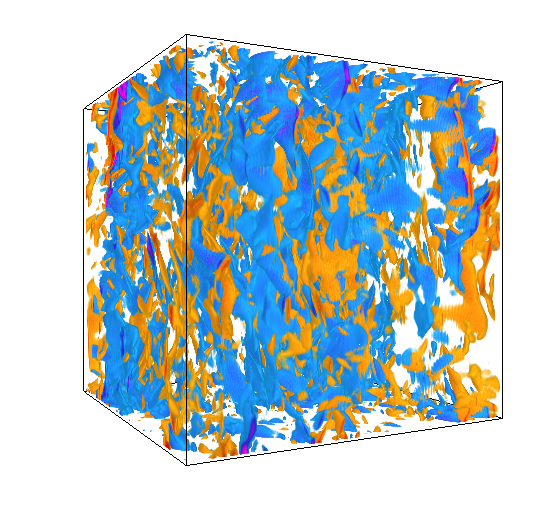}\\
        \hspace{-1cm} Top view $\theta$ \hspace{5.0cm} Top view $U$\\
    \includegraphics[width=0.49\textwidth,height=0.45\textwidth]{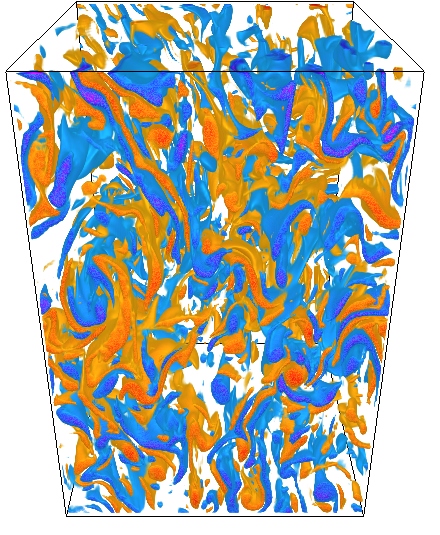}
    \includegraphics[width=0.49\textwidth,height=0.45\textwidth]{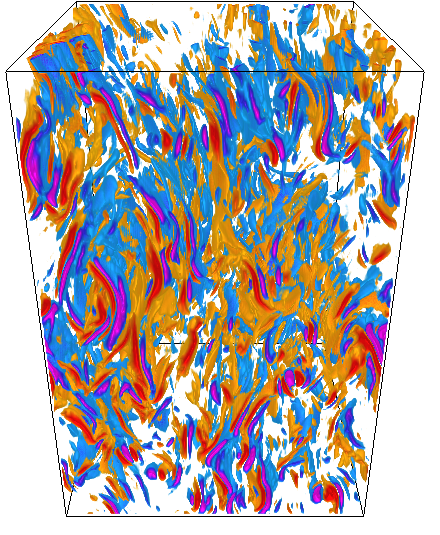}\\
    \caption{Snapshots of the temperature fluctuation $\theta$ (left column) and the ageostrophic velocity component $U$ (right column) from run D12 (with $\ekman=10^{-15}$, $\widetilde{Ra}=80$, $\Pr=1$) in the steady state, where a saturated LSV is present. The orientation in all panels is identical to Fig. \ref{fig:vis_press_omz}. Top row: side view. Bottom row: same data as in the top row, viewed from top.  Blue color indicates negative values while orange and red colors indicate positive values.}
\label{fig:vis_temp_Uag}
\end{figure}

Here, we provide some visualizations of the various fields from run D12, at the low Ekman number of $\ekman=10^{-15}$ with $Pr=1$ and $\widetilde{Ra}=80$. This run is in the geostrophic turbulence regime and the visualizations are produced in the statistically steady state, where a pronounced large-scale vortex (LSV) is present. The software Vapor \cite{li2019vapor} was used to generate the visualizations. We indicate positive values by orange and red contours and negative values by light and dark blue contours. 

The left column of Fig.~\ref{fig:vis_press_omz} shows the pressure field, where the large-scale columnar vortices are clearly visible. The right column of Fig.~\ref{fig:vis_press_omz} shows the $y$-component $v$ of the velocity, which displays smaller-scale features than the pressure field. Figure~\ref{fig:vis_velx_velz} shows the vertical vorticity $\omega_z$ (left column) and the vertical component $w$ of the velocity (right column). In the $u$ field, a clear trace of the large-scale vortex column is visible, while this is less obvious in the $w$ snapshot. Finally, Fig.~\ref{fig:vis_temp_Uag} shows the temperature perturbation field $\theta$ (left column) and the ageostrophic $x$ velocity $U$. Both fields show small-scale structures without any visible trace of the large-scale vortex.

\section{Conclusions}
\label{sec:conclusions}
In this work, we introduced the \avk{\textit{Rescaled Rapidly Rotating incompressible Navier-Stokes Equations} referred to as RRRiNSE} -- a new formulation of the Navier-Stokes equations in the Boussinesq approximation describing rotating Rayleigh-B\'enard convection, informed by the scalings valid in the asymptotic limit  $\ekman\to0$. We solved these equations for stress-free boundary conditions using the quasi-inverse method to perform efficient DNS in a previously unattainable parameter regime of extremely small Ekman numbers $\ekman$ relevant to geophysical and astrophysical fluid flows. We showed that the reduced equations of motion derive their increased efficiency from being well conditioned, thereby eliminating spurious growing modes that otherwise lead to numerical instabilities at small $\ekman$. We have validated our simulation results against published results in the literature, and showed that the vertical diffusion terms can be treated either implicitly or explicitly for small $\ekman$ owing to their smallness. We demonstrated for the first time that the full DNS of the \avk{RRRiNSE} converge to the asymptotically reduced equations for small $\ekman$, and showed that the time derivative in the mean temperature is inconsequential for the accurate determination of the average Nusselt number in the statistically stationary state, thus allowing a reduction by orders of magnitude in the simulation time required.

The results presented here provide an important advance in the numerical treatment of rotating convection in the rapid rotation regime, which will make it possible to explore for the first time the previously unattainable parameter regime of small but finite Ekman and Rossby numbers. \avk{The first study applying the approach described here to this regime, exploring the flow physics found there, is given in \cite{van2024bridging}.} Future studies will address the physics of the transition to the asymptotic parameter regime and the properties of optimal heat transport in rapidly rotating convection and the associated Reynolds numbers. Another direction for future investigation concerns the possibility of misalignment between the rotation axis and gravity, which has previously been studied in the context of the tilted $f$-plane using both the asymptotically reduced equations of motion \cite{kJ98,julien_ellison_knobloch_jfm} and the full rotating Boussinesq equations \citep{Novi2019,barker2020,cai2020penetrative}.

\section*{Acknowledgements}
K. J. passed away before this manuscript was finalized. We have attempted to present the results of our collaboration in accordance with his high standards. Any errors or misinterpretations remain our own. This work was supported by the National Science Foundation (Grants DMS-2009319 and DMS-2308338 (KJ), Grants DMS-2009563 and DMS-2308337 (EK)), by the German Research Foundation (DFG Projektnummer: 522026592, AvK) and by Agence Nationale de le Recherche (Grant ANR-23-CE30-0016-01, BM). This research used the Savio computational cluster resource provided by the Berkeley Research Computing program at the University of California Berkeley (supported by the UC Berkeley Chancellor, Vice Chancellor for Research, and Chief Information Officer). This research also utilized the Alpine high performance computing resource at the University of Colorado Boulder. Alpine is jointly funded by the University of Colorado Boulder, the University of Colorado Anschutz, and Colorado State University. Data storage for this project was supported by the University of Colorado Boulder PetaLibrary. This project also made use of computational resources from TGCC provided by GENCI (2024-A0162A10803).

\section*{Declaration of interests}
The authors declare no conflict of interests.

\appendix
\section{Mixed vorticity-velocity formulation of the \avk{RRRiNSE}}
\label{sec:app_mixed_vel_vort_formulation_RiNSE}
The primitive variable formulation of the \avk{RRRiNSE} in terms of $\ub=(u,v,w)$, $\boldsymbol{U}_\perp=(U,V)$, $\pi$, and $\overline{\Theta},\theta$ given by equations \eqref{eq:rinse_tot} 
in the main text is of $11^{th}$ order in $Z$. Specifically, the continuity equation requires, e.g., 
the imposition of an $11^{th}$ auxiliary boundary condition applied to the pressure function. Instead of pursuing this option, we numerically solve the following modified set of equations for the variables $\ub=(u,v,w)$, $\boldsymbol{\omega}=(\omega_x,\omega_y,\omega_z )$, $\boldsymbol{U}_\perp=(U,V)$, $\pi$ and $\overline{\Theta},\theta$:
\begin{subequations}
\label{eq:rinse_mixedVelocityVorticity}
\begin{align}
U = \frac{1}{\epsilon} (u+\partial_y {\pi}), \hspace{1cm} V = \frac{1}{\epsilon}(v - \partial_x {\pi}), \hspace{1cm}\omega_z = \partial_x v - \partial_y u,  
\label{eq:rinse_defs}
\end{align}
\begin{align}
\epsilon \partial_Z v - \partial_y w  + \omega_x \ = \ & 0,\\
\epsilon \partial_Z u - \omega_y - \partial_x w \ = \ & 0, \\
\partial_x U  + \partial_y V  + \partial_Z w \ = \ & 0, \\
\partial_t u - V - \mathcal{D}_u \ = \ &- \mathcal{N}_u, \label{eq:rinse_umom}\\
\partial_t v + U - \mathcal{D}_v \ = \ & - \mathcal{N}_v, \\
\partial_t w + \partial_Z {\pi}  - \frac{\widetilde{Ra}}{\Pr} \theta - \mathcal{D}_w \ = \ & -\mathcal{N}_w, \\
\partial_t \theta + (\partial_Z \overline{\Theta}-1) w - \mathcal{D}_\theta \ = \ & - \mathcal{N}_\theta,\\
\epsilon^{-2} \dst \overline{\Theta} - \mathcal{D}_\Theta = &\ - \mathcal{N}_\Theta,
\label{eq:rinse_Theta_mean}
\end{align}
\end{subequations}
where the linear diffusion and nonlinear advection terms are given by
\begin{align*}
- \mathcal{D}_u =&\ \partial_y \omega_z - \epsilon\partial_Z \omega_y, \hspace{0.5cm} & -\mathcal{N}_u = & \ \omega_z v - \omega_y w,\\
-\mathcal{D}_v =&\ \epsilon \partial_Z \omega_x - \partial_x \omega_z, \hspace{0.5cm}  & -\mathcal{N}_v  = &\ \omega_xw - \omega_z u, \\
-\mathcal{D}_w =&\ \partial_x \omega_y - \partial_y \omega_x,  \hspace{0.5cm}& -\mathcal{N}_w  = & \ \omega_y u - \omega_x v,\\
-\mathcal{D}_ \theta =&\ - \frac{1}{\Pr} (\partial_x^2 + \partial_y^2 + \epsilon^2 \partial_Z^2), \hspace{0.5cm}&  -\mathcal{D}_\theta = &\  - \partial_x (u\theta) - \partial_y (v\theta) - \epsilon \partial_Z (w\theta), \\
-\mathcal{D}_ \Theta =&\ - \frac{1}{\Pr} \partial_Z^2, \hspace{0.5cm}&  -\mathcal{N}_\Theta = &\   -  \partial_Z \overline{(w\theta)}.
\end{align*}
The above equations are of $10th$ order in $Z$ and do not require an auxiliary pressure boundary condition. We apply impenetrable, stress-free, fixed-temperature boundary conditions at the top and bottom which provide the $10$ conditions
\begin{equation}
w = \omega_x = \omega_y = \theta = \overline{\Theta} =  0  \text{ at } Z = 0,1.
\end{equation}
An immediate consequence of this formulation is  the fact that $\partial_Z p=0$ on $Z=0,1$. 
We note that the CFL constraints on the linear and nonlinear terms are identical to those presented in Table~\ref{Table:CFL}.

\section{The quasi-inverse method with Chebyshev-Galerkin bases}
\label{appendix:coral}
We illustrate how the direction $z$ is treated in \coral using the quasi-inverse method and Galerkin bases. This technique can readily be applied to coupled sets of equations of arbitrary order. For brevity and clarity, however, we consider the simple case of the second-order, scalar heat equation:
\be
\partial_t \phi - (\partial_{zz}-k_\perp^2) \phi = b(z) \label{eq:heat1d}
\ee
on the interval $z\in[-1,1]$, where the right-hand side $b$ contains explicit contributions (e.g. source terms or advection). 

We suppose this second-order equation is supplemented with two linear and homogeneous boundary conditions, a common case in fluid mechanics. By computing linear combinations of $N$ Chebyshev polynomials $\lb T_n \rb _{0\le n < N}$, one defines a Galerkin family of function $\lb\Phi_m \rb_{2,N}$:
\be 
\Phi_m (z)= \sum_{0\le n <N} S_{mn}T_n(z)\,, \label{eq:galerkinBasis}
\ee
each of which obeys the linear, homogeneous boundary conditions. It is crucial to note here that, as a result of enforcing these two boundary conditions, the Galerkin basis has been reduced as compared with the initial Chebyshev basis and now contains only $N-2$ polynomials. Next, we expand the variable $\phi$ in this Galerkin basis:
\begin{equation}
\phi(z,t) = \sum_{2\le m < N} \widetilde{\phi_m}(t) \Phi_m (z)\,.\label{eq:galerkinExpansion}
\end{equation}
The standard discretization of Eq.~(\ref{eq:heat1d}) consists in using expansion~(\ref{eq:galerkinExpansion}) and projecting on Chebyshev polynomials. Owing to the presence of derivatives $\partial_z$, this is conducive to dense (triangular) and, perhaps more importantly, ill-conditioned matrices \cite{boydBOOK}. The spirit of the quasi-inverse method consists in integrating the differential equation repeatedly, until the reformulated problem is clear of derivatives. In our case, we integrate Eq.~(\ref{eq:heat1d}) with respect to $z$ twice:
\be
\partial_t \iint \phi - \left(1 - \iint k_\perp^2\right) \phi = \iint b(z) + a_0 + a_1 z\,, \label{eq:heat1dQI}
\ee
where $a_0$ and $a_1$ are two arbitrary integration constants. Fortunately, these unknown constants appear in (and pollute) the $T_0(z)$ (constant) and $T_1(z)$ (linear) projections only. By projecting Eq.~(\ref{eq:heat1dQI}) on the unpolluted $N-2$ higher Chebyshev polynomials, one obtains an algebraic system for the $N-2$ unknown Galerkin coefficients $\widetilde{\phi}_m$. Denoting the natural scalar product for Chebyshev polynomials with brackets, $\left\langle ... \right\rangle$, we have for $2\le m,p <N$:
\begin{multline}
\sum_{0\le n <N} S_{mn}\left\langle T_p(z), \iint  T_n(z)\right\rangle \partial_t \widetilde{\phi_m} - S_{mp}\widetilde{\phi_m} \\+ k_\perp^2 \sum_{0\le n <N} S_{mn}\left\langle T_p(z), \iint T_n(z)\right\rangle \widetilde{\phi_m} = \left\langle T_p(z), \iint b(z)\right\rangle \,,\label{eq:heat_algebraic}
\end{multline}
where we have used the orthonormality condition $\left\langle T_p(z), T_n(z)\right\rangle = \delta_{pn}$. Crucially, the matrix representing the double integration,
\be
\left\langle T_p(z), \iint  T_n(z)\right\rangle \,,
\ee 
is penta-diagonal and, more importantly, well-conditioned. 

Finally, a discussion is in order considering the Galerkin stencil $S_{mn}$. Some care must be taken when defining the Galerkin basis, among all the possibilities. Considering the simple case of Dirichlet boundary conditions on both boundaries, one may be tempted by the following simple recombination:
\be
\Phi_{2p}(z) = T_{2p}(z) - T_0(z)\quad \mathrm{and}\quad \Phi_{2p+1}(z) = T_{2p+1}(z) - T_1(z)\,.
\ee
However, a dense discretization would result and therefore this stencil should be avoided. Instead, one should use
\begin{equation}
\Phi_m(z) = T_m(z) - T_{m-2}(z)\,,
\end{equation}
which is a well-conditioned and banded stencil. Thus, all coupling matrices appearing in Eq.~(\ref{eq:heat_algebraic}) are also banded and the system can be efficiently marched in time implicitly, e.g. with a Runge-Kutta scheme.

We emphasize that this procedure, exemplified on a simple scalar equation, can be generalized to systems of coupled PDEs without noticeable difficulty (but at the cost of increased book-keeping), as long as linear and homogeneous boundary conditions are imposed.

\section{Implicit-explicit time discretizations and the quasi-inverse method}
\label{sec:IMEX}
In this appendix, we summarize the specific formulation of implicit-explicit time discretization for both the NHQG model (\ref{eqns:reduced}) and the rescaled equations (\ref{eq:rinse_tot}) and (\ref{eq:rinse_mixedVelocityVorticity}). In all generality, these governing equations are represented by a system of the form:
\be
\label{eq:tstepA}
\lb \partial_t \mathcal{M} - \mathcal{L}_I \rb \vb^{(n+1)} =  \mathcal{L}_E \vb^{(n)} + \mathcal{N} (\vb^{(n)},\vb^{(n)}) +\boldsymbol{\mathcal{F}}^{(n)}_\theta.
\ee
where the superscripts $(n+1)$ and $(n)$ denote implicit (unknown) and explicit (known) variables from a prior time step, and $\vb$ represents the dependent variables associated with the hydrodynamics problem only. Generalization to the full problem including thermal effects (\ref{eq:tstep}) is straightforward and omitted for brevity. We summarize below the expressions for the various operators appearing in this equation for the different equations studied here.

\begin{itemize}
\item[A.] \underline{NHQG-RRBC:}  $\vb=(\Psi\; , \; w)^T$ and differential operators for equations (\ref{eqns:reduced}a,b)
\bea
\mathcal{M} =
\left (
\begin{array}{c|c}
\nabla^2_\perp & 0\\ \hline
0 & 1
\end{array}
\right ),\quad
\mathcal{L}_I =
\left (
\begin{array}{c|c}
\nabla^4_\perp  & \partial_Z \\
\hline
-\partial_Z & \nabla^2_\perp 
\end{array}
\right ),\quad
\mathcal{L}_E = \boldsymbol{0}_2,\quad
\mathcal{N} =
\left (
\begin{array}{c|c}
J[\Psi,\ ]  & 0 \\
\hline
0 & J[\Psi,\ ]
\end{array}
\right ).\ 
\eea
\begin{figure}
\label{fig:spy}
\centering
\includegraphics[width=0.32\textwidth]{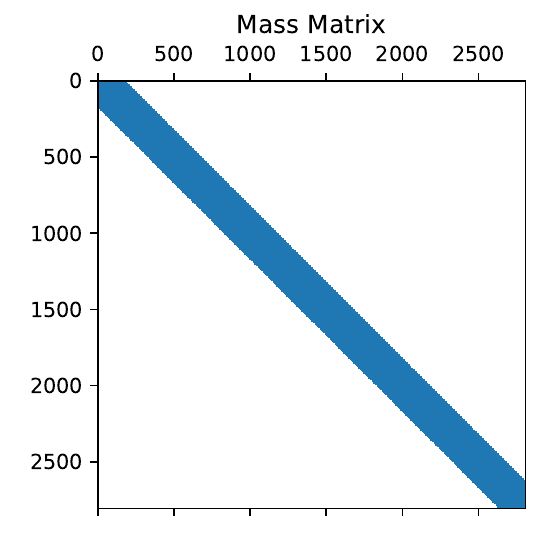}
\includegraphics[width=0.32\textwidth]{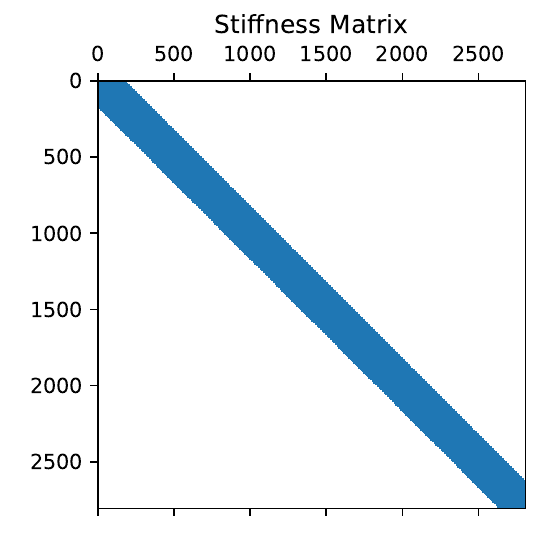}
\includegraphics[width=0.32\textwidth]{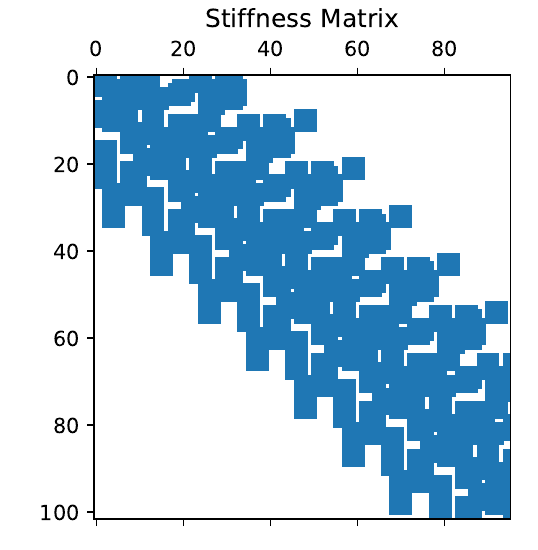}
\caption{\label{fig:spy_matrices} Spy plots of the mass $\mathcal{M}$ (left) and stiffness $\mathcal{L}_I$ (center) matrices, both in banded format. The fine structure of the stiffness matrix is illustrated in the close-up in the right panel.}
\end{figure}
\item[B.] \underline{RiNSE: Primitive variable formulation}:  $\vb^{(j)}=(\ub \; , \; {\boldsymbol{U}}_\perp, \pi)$ for equations (\ref{eq:rinse_tot}a-d) and
\bea
\mathcal{M} =
\left (
\begin{array}{c|c}
\boldsymbol{\mathcal{I}}_3 & \boldsymbol{0}_3\\
\hline
\boldsymbol{0}_3 & \boldsymbol{0}_3
\end{array}
\right ),\quad
\mathcal{L}_I =
\left (
\begin{array}{c|c}
-\nabla^2_\perp \boldsymbol{\mathcal{I}}_3 & \left ( \begin{array}{c|c} \boldsymbol{\mathcal{J}}_2 & 0  \\ \hline 0 \ \ 0  & \partial_Z  \end{array}\right )\\
\hline
\left ( \begin{array}{c|c}   \boldsymbol{\mathcal{J}}_2 & 0  \\ \hline 0 \ \ 0  & \partial_Z  \end{array}\right )  & 
\left ( \begin{array}{c|c}   -\epsilon\boldsymbol{\mathcal{J}}_2 &  \nabla_\perp^{T} \\ \hline  \nabla_\perp & 0  \end{array}\right ) 
\end{array}
\right ),\quad
\eea
\bea
\mathcal{L}_E =
\epsilon^2 \left (
\begin{array}{c|c}
\boldsymbol{\mathcal{I}}_3 \partial^2_Z  & \boldsymbol{0}_3\\
\hline
\boldsymbol{0}_3 & \boldsymbol{0}_3
\end{array}
\right ),\quad
\mathcal{N} = -
\left (
\begin{array}{c|c}
 \left (\ub_\perp\cdot\nabla_\perp +\epsilon w \partial_Z\right) \boldsymbol{\mathcal{I}}_3 & \boldsymbol{0}_3\\
\hline
\boldsymbol{0}_3 & \boldsymbol{0}_3
\end{array}
\right ).
\eea
Here $\boldsymbol{\mathcal{I}}_3$ is the order three identity matrix and 
\bea
\boldsymbol{\mathcal{J}}_2 =
\left ( 
\begin{array}{cr}
0  & -1 \\ 1 & 0 
\end{array} \right ).
\eea
Figure~\ref{fig:spy_matrices} demonstrates the sparsity of the quasi-inverse approach via spyplots for the mass and stiffness matrices $\mathcal{M}$ and $\mathcal{L}_I$, respectively.
\item[C.] \underline{RiNSE: Mixed vorticity-velocity formulation}:  $\vb^{(j)}=(\ub \;\vert \;{\boldsymbol{U}}_\perp, \pi \;\vert \; \boldsymbol{\omega})$ (see \ref{sec:app_mixed_vel_vort_formulation_RiNSE}) and
\bea
\mathcal{M} =
\left (
\begin{array}{c|c|c}
\boldsymbol{\mathcal{I}}_3 & \boldsymbol{0}_3 & \boldsymbol{0}_3\\
\hline
\boldsymbol{0}_3 & \boldsymbol{0}_3  & \boldsymbol{0}_3\\
\hline
\boldsymbol{0}_3 & \boldsymbol{0}_3  & \boldsymbol{0}_3
\end{array}
\right ),\quad
-\mathcal{L}_E = \epsilon
\left (
\begin{array}{c|c|c}
\boldsymbol{0}_3 &\boldsymbol{0}_3 &
\left( \begin{array}{c|c}  \partial_Z\boldsymbol{\mathcal{J}}_2 &  0 \\ \hline  0 & 0 \end{array}\right ) 
\\
\hline
\boldsymbol{0}_3 & 
\boldsymbol{0}_3 &
\boldsymbol{0}_3\\
\hline
\left( \begin{array}{c|c}  \partial_Z\boldsymbol{\mathcal{J}}_2 &  0 \\ \hline  0 & 0 \end{array}\right )
& \boldsymbol{0}_3 & \boldsymbol{0}_3
\end{array} 
\right ),
\eea
\bea
-\mathcal{L}_I = 
\left (
\begin{array}{c|c|c}
\boldsymbol{0}_3 & \left ( \begin{array}{c|c} \boldsymbol{\mathcal{J}}_2 & 0  \\ \hline  0  & \partial_Z \end{array}\right ) &
\left( \begin{array}{c|c} 0 &  -\nabla^{\perp T} \\ \hline  \nabla^\perp & 0 \end{array}\right ) 
\\
\hline
\left ( \begin{array}{c|c}  \boldsymbol{\mathcal{J}}_2 & 0  \\ \hline 0   & \partial_Z  \end{array}\right )  & 
\left ( \begin{array}{c|c}   -\epsilon \boldsymbol{\mathcal{J}}_2 &    \nabla_{\perp}^ T \\ \hline  \nabla_\perp & 0  \end{array} \right ) &
\boldsymbol{0}_3\\
\hline
\left( \begin{array}{c|c} 0 &  -\nabla^{\perp T} \\ \hline  \nabla^\perp & 0 \end{array}\right )
& \boldsymbol{0}_3 & \boldsymbol{\mathcal{I}}_3
\end{array} 
\right ),
\eea
and 
$\mathcal{N}=Diag\left[\{\mathcal{N}_u,\mathcal{N}_v,\mathcal{N}_w\},\boldsymbol{0}_3,\boldsymbol{0}_3\right ]$.
\end{itemize}
From $\mathcal{L}_I$ we see that the system \eqref{eq:tstepA} is in each case of second order in $Z$ requiring impenetrable boundary conditions $w=0$ at $Z=0,1$. For cases B and C stress-free boundary conditions ($\partial_Z \ub_\perp =0$) 
are enforced via an appropriate Chebyshev-Galerkin basis $\Phi_j(Z)$ for each variable \textcolor{black}{(no-slip boundary conditions, $\ub_\perp =0$, can also be considered in this approach, but are not analyzed here)}. The above holds regardless of whether $\mathcal{L}_E$ is treated explicitly or implicitly. For case B only, an implicit treatment of $\mathcal{L}_E$ increases the order of the system to seven and an additional auxillary boundary condition on the pressure is required. This does not occur for Case C which is preferred.

\section{Analysis of the mean temperature equation}
\label{sec:tempomit}

The mean temperature equation
\be
\epsilon^{-2} \pd{t}  \overline{\Theta} + \pd{Z} \lb \overline{w \theta} -  \frac{1}{Pr}\pd{Z}  \overline{\Theta}  \rb = 0 
\ee
in a statistically stationary state implies
\be
  Nu_t - 1  = {Pr}\,   \overline{\overline{w \theta}}^t -   \overline{\pd{Z} \overline{\Theta}}^t\quad\implies\quad    
 Nu_t - 1 =  Pr \lbr  \overline{\overline{w \theta}}^t \rbr_{Z} .
\ee
With this interpretation $Nu_t$ is strictly a constant. It follows
\be
\epsilon^{-2} \pd{t}  \overline{\Theta} + \pd{Z} \lb \overline{w \theta} - \overline{\overline{w \theta}}^t -  \frac{1}{Pr}\lb \pd{Z}  \overline{\Theta} -   \overline{\pd{Z} \overline{\Theta}}^t  \rb 
\rb = 0 
\ee
and given  $\epsilon^2\overline{\Theta}_{2^+}\equiv\sum_{j\ge 2} \epsilon^j \overline{\Theta}_j$, $\pd{t}\overline{\Theta}_{(0,1)} =0 $ such that 
$\pd{Z}\overline{\Theta}_{(0,1)} = \pd{Z}\overline{\overline{\Theta}}^t_{(0,1)}$. This implies 
\be
 \pd{t}  \overline{\Theta}_{2^+} + \pd{Z} \lb \overline{w \theta} - \overline{\overline{w \theta}}^t -  \frac{\epsilon^2}{Pr}\lb \pd{Z}  \overline{\Theta}_{2^+} -   \overline{\pd{Z} \overline{\Theta}}_{2^+}^t  \rb 
 \rb = 0 .
\ee
To leading order 
\be
 \pd{t}  \overline{\Theta}_{2} + \pd{Z} \lb \overline{w \theta} - \overline{\overline{w \theta}}^t  
 \rb \approx 0 
\ee
indicating that $\mathcal{O}(1)$ fluctuations in the heat transport about the mean $Nu_t$ are accounted for by mean temporal variations in the mean temperature at $\mathcal{O}(\epsilon^2)$, i.e., 
$\overline{\Theta}_2$.
However, for numerical efficiency it is found that the temporal fluctuations of $\epsilon^{-2} \pd{t}  \overline{\Theta}$ can be neglected. Hence, 
\be
 \pd{Z} \lb \overline{w \theta} -  \frac{1}{Pr}\pd{Z}  \overline{\Theta}  \rb = 0 \,,
\ee
resulting in the time-dependent Nusselt number  
\be
 Nu(t) - 1  =Pr\,  \overline{w   \theta}   - \pd{Z}  \overline{\Theta} .
\ee
This implies 
\be
  Nu(t) - 1 =  Pr \lbr  \overline{w \theta}   \rbr_Z  \quad\implies\quad  \overline{Nu(t)}^t - 1 =  Pr \overline{\lbr  \overline{w \theta}\rbr_Z}^t.
\ee
The difference in averaged Nusselt numbers from the two methods is given by
\be
  \overline{Nu(t)}^t - Nu_t =  Pr \lb \overline{\lbr  \overline{w \theta}\rbr_Z}^t - \lbr  \overline{\overline{w \theta}}^t \rbr_Z\rb
  = \left . \lb \pd{Z}  \overline{\Theta} -   \overline{\pd{Z}\overline{\Theta}}^t\rb \right \vert_{0,1} .
\ee
If the operations of depth- and time-averaging commute then assuming equivalence in the thermal and velocity mean statistics implies  $\overline{Nu(t)}^t = Nu_t$. 

Moreover, if $\epsilon\mapsto\epsilon^*$ such that $\epsilon^* > \epsilon$ then it follows
\be
  Nu^*_t - Nu_t =  Pr \lb  \lbr  \overline{\overline{w \theta}}^{*t} \rbr_Z - \lbr  \overline{\overline{w \theta}}^t \rbr_Z \rb
  = \left . \lb \overline{\pd{Z}\overline{\Theta}}^{*t} -   \overline{\pd{Z}\overline{\Theta}}^t\rb \right \vert_{0,1} .
\ee
If the period of time-averaging is sufficiently long then assuming equivalence in the thermal and velocity mean statistics this implies  $Nu^*_t = Nu_t$. 

It is found that the magnitudes $\vert \overline{Nu(t)}^t - Nu_t\vert$ or $\vert Nu^*_t - Nu_t\vert$ depend on the horizontal domain size upon which area-averaging is performed. Increasingly larger domains contain greater statistical sampling advantageous to the aforementioned commutation that results in asymptotic error convergence.

\avk{\section{Comparison of iNSE and RRRiNSE with and without slaving \label{app:comparison_slaving_rescaling}}}
\avk{
Here we present results on different variations of the numerical approach presented in the main text. Specifically, we compare simulations of the iNSE and of the RRRiNSE with and without slaving of the mean temperature field. For the tests described below, we fix $\widetilde{Ra}=40$, while setting $N_x=N_y=128$, $L_x=L_y=10 \ell_c$, and we employ the second order ARS-$222$ time-stepping scheme. We set the Ekman number     to low values, either $Ek=10^{-9}$ or $Ek=10^{-12}$. We have tested different values for the vertical resolution $N_Z$, since an increased $N_Z$ is seen in Fig.~\ref{fig:Param_Re_vs_Ek} to lead to less accurate spectra in the unscaled equations.}

\begin{figure}
    \centering
    \includegraphics[width=0.482\linewidth]{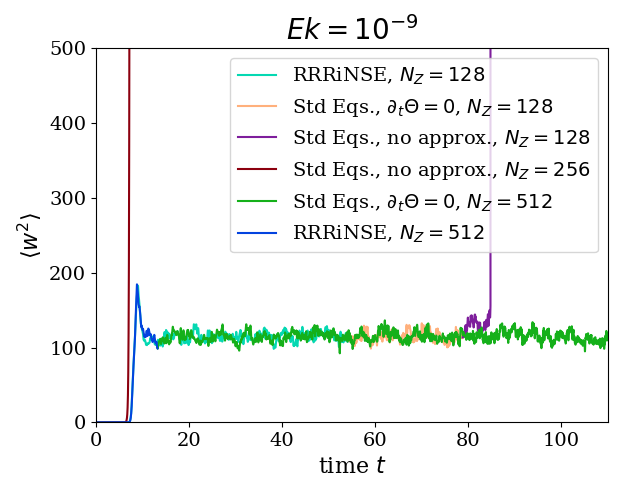} \includegraphics[width=0.465\linewidth]{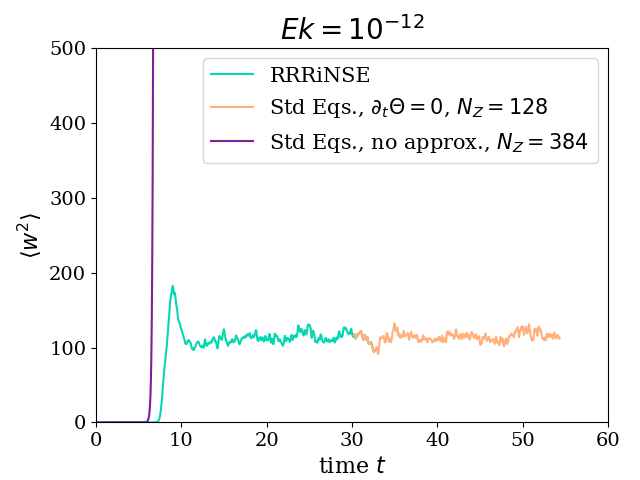}
    \includegraphics[width=0.482\linewidth]{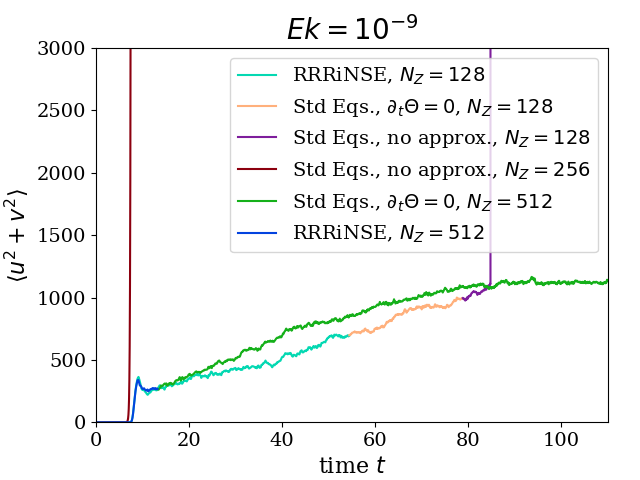}
    \includegraphics[width=0.482\linewidth]{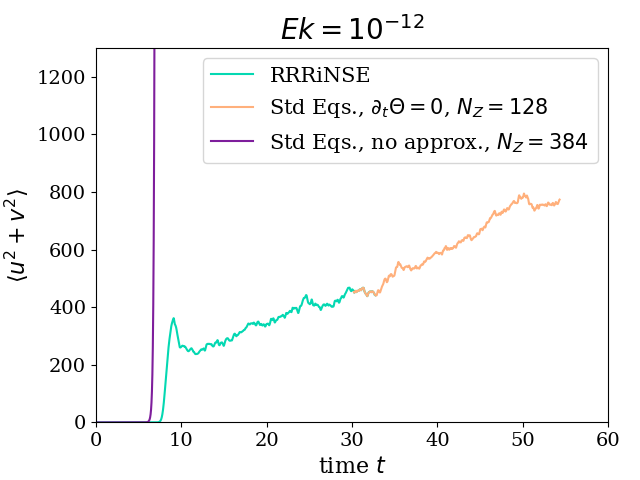}
    \caption{Comparison of the RRRiNSE with the unscaled equations including ({\it no approx.} in legend) or omitting ($\partial_t \Theta=0$ in legend) the slow mean temperature time derivative, showing time series of  vertical kinetic energy (top row) and horizontal kinetic energy (bottom row). All quantities are plotted in rescaled units, regardless of the dimensionless formulation adopted for the governing equations. The horizontal kinetic energy grows as a consequence of inverse energy cascade.}
    \label{fig:comparison_stability}
\end{figure}

\avk{The results of these tests are summarized in Fig.~\ref{fig:comparison_stability} showing the time series of the kinetic energy in horizontal (top row) and vertical (bottom row) motions from different simulations. First, we note that the RRRiNSE are found to be unconditionally stable in all cases. In contrast, the standard iNSE are unstable when a slow adjustment of the vertical profile is permitted. This is seen to be the case when {small amplitude, random perturbations are used for initial conditions at} both $Ek=10^{-9}$ (crimson line in the left column of Fig.~\ref{fig:comparison_stability}) and $Ek=10^{-12}$ (purple line in the right column of Fig.~\ref{fig:comparison_stability}). In addition to using noise as initial condition, we have also considered an equilibrated RRRiNSE solution, which was suitably rescaled and used as an initial condition for continued time-marching of the solution with the unscaled iNSE. This procedure yielded the following observations: including the slow mean temperature time derivative (e.g. purple line on the left panels) is associated with eventual blow-up. 
Rather surprisingly, despite the presence of spurious modes polluting the numerical spectra, we have not observed a numerical blow-up when starting from an equilibrated solution and simultaneously omitting the slow mean temperature time derivative (e.g. salmon lines on both panels). When all terms in the mean temperature equation are retained, including the slow time derivative of the mean temperature, the unscaled equations quickly diverge (see the red and purple lines in the left panel). When the mean temperature derivative is set to zero, the unscaled equations of motion remain surprisingly stable for the times investigated, despite the {presence of spurious growth rate values in the spectra obtained when solving the numerically ill-conditioned generalized eigenproblem}.}
\bibliography{jcp_bib}

\end{document}